\documentclass[epjH]{svjour}
\usepackage[most]{tcolorbox}
\usepackage{enumitem}
\usepackage{lipsum}

\usepackage{graphicx}
\usepackage{hyperref}

\begin{document}
\title{The Charm of Theoretical Physics (1958-1993)\footnote{The text presented here has been revised by the authors based on the original oral history interview conducted by Luisa Bonolis and recorded in Rome, Italy, 1--3 March 2016.}
\subtitle{Oral History Interview}}

\author{
Luciano Maiani\inst{1}\fnmsep\thanks{\email{luciano.maiani@roma1.infn.it}}
\and Luisa Bonolis\inst{2}\fnmsep\thanks{\email{lbonolis@mpiwg-berlin.mpg.de, luisa.bonolis@roma1.infn.it}}
}
\institute{Dipartimento di Fisica and INFN, Piazzale A. Moro 5, 00185 Rome, Italy \and  Max Planck Institute for the History of Science, Boltzmannstra\ss e 22, 14195 Berlin, Germany}

\abstract{Personal recollections on theoretical particle physics in the years when the Standard Theory was formed. In the background, the remarkable development of Italian theoretical physics in the second part of the last century,  with great personalities like Bruno Touschek, Raoul Gatto, Nicola Cabibbo and their schools. 
} 
\maketitle
\tableofcontents

\section{Apprenticeship}


 L. B. \hspace{0.2 cm}
How did your interest in physics arise? You enrolled in the late 1950s, when the period of post-war reconstruction of physics in Europe was coming to an end, and Italy was entering into a phase of great expansion. Those were very exciting years. It was the beginning of the space era\dots
\vskip 0.3 cm

\noindent L. M. \hspace{0.2 cm}
The beginning of the space era certainly had a strong influence on many people, absolutely. The landing on the moon in 1969 was for sure unforgettable, but at that time I was already working in Physics and about to get married\dots \ My interest in physics started well before.  The real beginning was around 1955. Most important for me was astronomy. It is not surprising that astronomy marked for many people the beginning of their  interest in science. 

When I was in the high-school, the lyceum, I was lucky to have as schoolmate Giuseppe Grimaldi, a very interesting and determined guy. Grimaldi later on became a priest and I was so grateful to him as to want him,  in '69, to marry us\dots \ At school Giuseppe was not brilliant --- the guy that doesn't speak much --- and I was taking much better votes than him, but he initiated me to astronomy. He had a telescope. So we were learning constellations,  studying the stars, so much that, at a certain point, I decided to build myself a telescope with which I could observe sunspots and other objects in the sky. But, most of all, in studying astronomy I discovered physics. 

There was a book by Giorgio Abetti, entitled \textit{The Sun} \cite{Abetti1952} which really fascinated me. With the sun, you discover Planck's law, spectroscopy, quantum mechanics\dots \ and that meant discovering the world. So I started by myself reading books that were available, but later I tried to go a little further than simply popular science books. I remember that I bought a volume published by Hoepli --- a publisher that was a good source of such books --- which tried to explain quantum mechanics. Of course it was not really teaching you quantum mechanics, but explaining what Sommerfeld did, what Pauli did\dots and it introduced the principle of action. It mentioned things like the adiabatic invariants, relativistic corrections to the Bohr formula\dots \ so it was not at all trivial.  I was totally fascinated and I decided that I had to study relativity and quantum mechanics and that this was what I wanted to get into. And then, I arrived at the end of high-school. It was a very quiet thing, I had  no problem in doing the maturity exam, essentially because I was studying 24 hours a day! I was secluded in my house, studying here and there, and I was very happy. At that time my elder sister was getting engaged with Giuseppe Signorelli --- whom she later married --- a chemist and a person that I liked very much.  He was doing research in industrial chemistry, he was a metallurgist who was  later called as full professor at the faculty of engineering in Rome. He gave me as a gift for my final high-school exam a famous book by Einstein and Infeld, \textit{The Evolution of Physics} \cite{EinsteinInfeld1938}. This was really the beginning. I really started getting excited!

While still at the high school, I had tried to get into particle physics and bought a book by Fermi,  \textit{Particle Physics} \cite{Fermi1951}, but I couldn't understand anything! The book spoke of waves, perturbation theory, and it contained many, many formulae. I still remember when I bought it, at a  book store near the university. I went home and read the very short introduction,  that was not very satisfactory, and then I got into topics that I could understand only several years  later, when I was teaching this kind of things to university students of the third year. At that point I had to quit. I still have the Einstein and Infeld book in my library, but I think I lost Fermi's one.

\vskip 0.3 cm

L. B. \hspace{0.2 cm}
However, from what you say, it appears that by that time you had fully discovered your passion for physics\dots
\vskip 0.3 cm

L. M. \hspace{0.2 cm}
Yes, indeed. And I also understood --- by reading the Hoepli book on quantum mechanics --- that I had to study mathematics, and so I studied it myself. The fianc\'ee of my sister gave me his classical university textbook by Aldo Ghizzetti. And I remember that, before the start of university courses, in fall 1959, after dining with my family, I used to retire in my room to study Ghizzetti's book. I came from classic lyceum, mostly dedicated to humanistic studies, but  when I went to the university I had already a good understanding of what mathematical analysis was. In addition, in Rome I had a wonderful teacher, Fernando Bertolini, who was alternating with Gaetano Fichera, a well known mathematician and an excellent teacher, too. 

Bertolini was so good in explaining mathematics, in a way that was simple and at the same time very deep. Not just making everything trivial but showing that you can have deep concepts and that you can explain them in a simple way. 
 \vskip 0.3 cm
 
L. B. \hspace{0.2 cm}
In this regard, there is a very nice quote from Einstein: ``You do not really understand something unless you can explain it to your grandmother!''. \\So you spent the first years studying and studying\dots
 \vskip 0.3 cm

L. M.  \hspace{0.2 cm}
And I was delighted! University fulfilled all my expectations! The point is that by that time I knew  many things --- quantum mechanics, relativity \dots --- but always in a way that you would not be able to explain to others, because such books tell you something, but to get a real basis you need to reconstruct the whole story. And university was the place where they would tell you the whole story. So I had no problems in general. But I had problems with chemistry\dots \ My personal interpretation is that they explained thermodynamics using ill defined concepts, meant to be ``intuitive'', so I could not understand what they were talking about. This is my understanding, but anyway I did not pay a lot of attention. In general, I had excellent grades. In chemistry I got 21/30, if I remember correctly, barely sufficient, but I was not discouraged by that. 
 \vskip 0.3 cm

L. B. \hspace{0.2 cm}Who were you teachers? And how did you organize your studies?
\vskip 0.3 cm

My plan was to study mathematics first. The official schedule of courses implied  to study ``General Physics'' at the same time, but I gave a look into the textbook by Gilberto Bernardini, \textit{General Physics}, and decided to postpone the exam in physics, because the book was not telling the story from the beginning and mathematics had to go first. So, I ended taking the exams of ``Physics I'' and ``Physics II'' with Edoardo Amaldi. He talked very well and was a very good teacher. With Amaldi, I completed the first two years. And then I came to the 3rd year and started with Marcello Cini, who was teaching ``Istituzioni di Fisica Teorica'' --- which later would be my first teaching in physics. Cini was a very intriguing person to me. He was not very good at explaining --- sometime he messed his arguments up --- but he confronted us with wonderful ideas and wonderful books, the real first contact with modern physics. In particular for Special Relativity, which I studied on Richard C. Tolman's book {\it Relativity, Thermodynamics and Cosmology}  \cite{Tolman1934}, a wonderful book I still consult from time to time, and then he gave us the book by Erwin Schr\"odinger \textit{Statistical Thermodynamics} \cite{Schroedinger1952} that is absolutely fantastic.\footnote{I still remember by hearth entire sentences from this book, like what Schr\"odinger writes after stating the Gibbs paradox in classical statistical physics: \textit{After a railway disaster, always authorities interrogate themselves how could that happen} \dots \ Or about undistinguishable quantum particles: \textit{Democritus of Abdera and not Max Planck has been the first quantum physicist!} } So Cini's course was really exciting. In the 4th year there was the course of ``Fisica Teorica'', taught by Enrico Persico, who had been Enrico Fermi's friend since school years. In 1926 both had  won the first competition for the new chairs of Theoretical Physics established in Italy. He had always been an exceptionally good teacher, but at that time maybe he was already too old, I did not find him very exciting. In fact I did not even go to all his lectures, because after following Cini's course I had encountered  Dirac's book \textit{The Principles of Quantum Mechanics} and I studied quantum mechanics by myself. In the evening, after dinner with my family, I always liked very much  to retire in my room to study. In the first two university years I took notes about analysis and mechanics, but after the 3rd year it was Dirac\dots \

Quietly reading in my room, Dirac's book was where I really got what I understand of quantum mechanics (even if I like to say, with Richard Feynman, that: ``Nobody understands quantum mechanics!'').  I got it first from  Dirac's book and later from Feynman's book on the theory of fundamental interactions \cite{Feynman1961}, which was for me the real introduction to elementary particle physics. 

To complete the picture of the books I loved while studying at the university, I must go a bit back in time, when I discovered the book by Hendrik A. Lorentz, \textit{The Theory of Electrons} \cite{Lorentz1909} published immediately after the appearance of Einstein's Theory of Special Relativity and only a few years after Max Planck's Quantum Theory. Lorentz's book presents an attempt to describe the properties of matter and radiation by applying Maxwell equations to the atomic electrons and it can be thus considered to be the last attempt of classical physics to get at the \textit{Theory of Everything}. I was completely fascinated by Lorentz's style of telling everything in words, in the text, and confine the equations in footnotes at the end, that I could study later at ease. This is still the way I read articles and books: first the words and then the formulae. I  remember reading systematically  Lorentz's  book on the bus while going from my place to the university and viceversa. 

Lorentz's book mapped the frontier where modern physics starts and where I wanted to get into. The honesty and the surprise with which he recognises that there were insurmuontable obstacles to the classical vision left in me a deep impression.

\vskip 0.3 cm

L. B. \hspace{0.2 cm}
Did you follow Bruno Touschek's course on statistical mechanics?
\vskip 0.3 cm

L. M.  \hspace{0.2 cm}
For us students, Bruno Toschek was a complete surprise. We, I at least, did not know him by fame, as was the case with Edoardo Amaldi or Gilberto Bernardini, but I was immediately impressed by his caliber of a world class  scientist. Edoardo Amaldi has written about his adventurous escape from nazism \cite{Amaldi 1981} and how he got to establish himself  in Rome after the war, and there are excellent biographies \cite{BonolisPancheri2011}, which tell what he was doing in the years we met him, namely building in Frascati an extraordinary accelerator of his conception, an electron-positron accumulation ring that was going to be the first of the family of modern colliders (I will be back on this later). 

However what struck us at that time, and I want to report here, was his quality as a teacher, the way he was introducing us to statistical mechanics using a modern language we had never heard in previous courses. He would present his lecture, consulting personal notes that he had probably prepared the night before, as if he had just discovered what he was illustrating. Extremely clear and precise, Touschek spoke a perfect Italian with a  fascinating Austrian accent and sometime old fashioned expressions. He referred to the heat bath to reach thermal equilibrium as ``vasca di bagno'' and described his revolutionary idea of making head-on electron-positron collisions as ``treno-contro-treno''. 

One could see the perfect image of a scientist and a perfect introduction to what research in physics might be. Meeting Bruno convinced me that research was my destination but, at the same time, made me doubt that I could be able to work in theoretical physics.

\vskip 0.3 cm

L. B. \hspace{0.2 cm}
What do you remember of your colleagues during those years? 
\vskip 0.3 cm

L. M.  \hspace{0.2 cm}
I had the good fortune to get into an extraordinary class, probably because the space adventure and the building of the electron synchrotron at Frascati had attracted to physics a lot of brilliant students. 
And so in my year there was a group in which very soon we got in touch with each other and, after the 3rd year, began to be recognized by the teachers. 

Sergio Doplicher was the \textit{wise man}.  I had studied excellent popular science books while being at the high-school, but apparently Doplicher  had studied Dirac's book while at the Gymnasium! So any time I had a doubt or any time I wanted to have an indication I would go to Sergio and ask: where can I study relativity? Or quantum mechanics? And so on. He always had the right reference. It was very funny that he himself was --- and still today is --- not very talkative and not very communicative, so at the exams, after all, both Guido Altarelli and me got better votes than Sergio. It was absurd because he was so advanced. Our interpretation was  that in fact he had problems in remembering older things that he had forgotten in the meantime! It would be the same if you asked me about topics I studied long ago. I could have problems in reconstructing\dots 

For some time I prepared exams  with Franco Buccella, very bright and competitive. Then there was Giovanni Gallavotti. Giovanni was also taken in great esteem in our group because it was clear that he was studying at a rate which was at least twice what we were doing. He was really excellent \dots You could ask him anything and he would tell you the right story. And there was, of course, Guido Altarelli, always flying very high: at the end he got the best votes of the whole class. 

I did not familiarise much with Guido during university years, our friendship came in Florence, and later in Rome, where we became strict collaborators. Giorgio Capon, a nephew of Enrico Fermi's wife Laura, was a very good friend of Guido, and he later became an experimental physicist in Frascati. I made good friendship with Giorgio Capon and  his wife Teresa quite later, in the late 1970s, while we were in Paris. 

There where others, in the group of the distinguished students, that went to experimental physics, and whom I did frequent. Among them Massimo Cerdonio, Piergiorgio Picozza, Piero Spillantini\dots \ Most people that are around now as full professors  come from that particular year, the year that started in '59/'60. I would like to mention other names: Pio Pistilli, Marcello Fontanesi, Claudio Procesi, Piero Monacelli, Guido Ciapetti\dots. So many! Professors, too, were very impressed by this group that was covering all parts of physics. 
\vskip 0.3 cm

L. B. \hspace{0.2 cm}
What were your feelings about your experience with laboratory courses and experimental physics? 
\vskip 0.3 cm

L. M.  \hspace{0.2 cm}
I took  the course on physics laboratory during the 4th year. One had to do one experiment of modern physics in a team of three students. I was with Massimo Cerdonio and Antonio de Gasperis.  I could see that Massimo was really a very good experimentalist and that I had not the mind of an experimental physicists. At that time we were in a way scared by theoretical physics; at least I was\dots  \ I learned later that Fermi used to say that if you are a theoretical physicist you must be very bright, while if you are an experimental physicist at least you can do useful things\dots .  This was the idea. Apparently Fermi had told something similar to Bruno Pontecorvo, when he examined him at his arrival in Rome from Pisa University in 1931.

\vskip 0.3 cm

L. B. \hspace{0.2 cm}
On the other hand Pontecorvo later turned out to be a very special physicist, making great contributions both to theory and to experimental physics. In this he was similar to Fermi. And in fact later Fermi rated  Pontecorvo as one of the most brilliant physicists he had ever met. And Fermi was generally not very prone to praise someone\dots \ 

Well, at this point I can imagine that you began to think to your dissertation\dots
\vskip 0.3 cm

L. M.  \hspace{0.2 cm}
Yes. Actually, when I was about to enter in my 4th year, I learned from friends that there were fellowships given by the physics laboratory of Istituto Superiore di Sanit\`a. I had high scores, so I thought: ``Why not?'' I went there and Mario Ageno, the director, said: ``I see you have many interests and so if you really want the fellowship, I will give you a fellowship. But you have to take it!'' . And so I took an experimental thesis with Giorgio Cortellessa, who was one of the senior physicists of the lab. That was the time when I  met Ugo Amaldi, the son of Edoardo, who was also working in that laboratory. In fact the presence of Ugo Amaldi was for me a very good indication that that had to be a good place, because he was obviously a very good researcher. There were also Giorgio Matthiae, with whom I have still ongoing collaborations, Gabriele Fronterotta, who shortly after moved to IBM, and many others\dots \ all excellent people. 

Being at Istituto Superiore di Sanit\`a was a good experience, except that I suffered a lot with the experimental thesis. Not because I did not like experimental physics; I like to work with instruments, I still do bricolage etc. But  first of all I was not so bright as Cerdonio, who really is able to understand how to do an experiment.  Secondly, there are a lot of practical things that go wrong and make you lose time. You lose time because things break down, because orders are not handled properly, because somebody steels your oscilloscope, etc.

Be as it may, I graduated in February '64, in the recuperation session of my 4th year. But Altarelli, Buccella and Gallavotti, who had taken a theoretical thesis with Raoul Gatto and with Bruno Touschek, had graduated within the regular year, in '63.  
Altarelli and Buccella, jointly, were working at a dissertation under the guidance of Gatto, computing the cross section of some electrodynamic processes which were useful to measure the luminosity of the machine then in construction in Frascati. It was a big electron positron collider that followed the pioneer work done by Touschek and collaborators with the accumulation ring AdA (Anello di Accumulazione) \cite{Bernardini2004,Bonolis2005,BonolisPancheri2011}. Gallavotti was also working on a similar process, under the guidance of Touschek. 

Being a larger version of AdA, the machine had been called Adone and it was the sensation of the moment in Rome. Big expectations were raised about the results to be obtained by what was the first exploration of Electrodynamics at high energy. Few years earlier, Raoul Gatto and Nicola Cabibbo had written a long article that summarised the theoretical situation of the high energy electron-positron collisions \cite{CabibboGatto1961b}. It was called {\it The Bible} by people in Frascati and showed very clearly the potential for elementary particle physics of the future experiments with Adone.

\vskip 0.3 cm

L. B.  \hspace{0.2 cm}
 Cabibbo and Gatto had inaugurated the theoretical discussion on possible experiments with electron-positron colliding beams with a preliminary pioneering article sent in February 1961 \cite{CabibboGatto1961a}. Nicola Cabibbo later recalled that while doing this work they had ``the exhilarating experience of expanding into a vacuum'' because for a few years the only theoretical papers on the physics of $e^+e^-$ were those issuing out of Rome or in Frascati  \cite[p. 221]{Cabibbo1997}.

What kind of work were you doing for your dissertation?

\vskip 0.3 cm

L. M.  \hspace{0.2 cm}
I was working on the new solid state, Silicon detectors. They were promising devices and indeed, later on, they turned out to play an extremely important role in experimental high energy physics. My task was limited, but interesting, namely to determine the energy resolution one could obtain with one such detector by measuring the energy of the gamma rays coming out of electron-positron annihilation in the semiconductor. 

I was very unhappy for the problems with my thesis work and, in addition, I felt isolated from the exciting developments in Frascati and started thinking to leave the field I had been put in at Sanit\`a, however promising it was. 

After a lot of thinking, in fall '63, I went to Ageno and told him: ``You are trying to address me to solid-state physics --- which was what I was doing --- and I am very grateful for all you are doing for me! However, this is not what I want to do. I want to study theoretical physics, I want to study quantum electrodynamics, I want to study particles. And so I am ready to give back my fellowship and try by myself.'' And to my surprise Ageno said: ``You want to be a theoretical physicist? Ok, you will be a theoretical physicist! Where do you want to go?''.  Of course, I could not answer precisely. Then Ageno asked Cini, who was acting as a consultant to the laboratory, whether Cini would take me. And Cini said that I was welcome to work with his group. However, at that time Altarelli and Buccella were already thinking to move to Florence where Raoul Gatto had been called to the chair of theoretical physics. Gallavotti, too, was going and so I decided to go to Florence. 

Meanwhile, many things had happened.

Buccella had come one day in Sanit\`a with the news that a new particle had been discovered, the $\Omega ^{-}$~~\cite{BarnesEtAl1964}, whose mass and properties had been predicted theoretically by Murray Gell-Mann \cite{GellMann1962} and by Susumu Okubo \cite{Okubo1962}.

Buccella brought with him a long article in \textit{Physics Today}~\cite{Weisskopf 1963} 
describing the recent developments in elementary particle physics. We sat down and studied the article and discussed for long, trying to get  something out of it. 

There were also news about the new theory put forward by Nicola Cabibbo \cite{Cabibbo1963}, on which I will return later, and the idea that protons, neutrons and the other subnuclear particles could be made by more elementary constituents, {\it quarks}, introduced by Murray Gell-Mann \cite{Gell-Mann1964} and by George Zweig~\cite{Zweig1964}; I did not know what it was but all looked very exciting. And so theoretical particle physics, more than Quantum Electrodynamics, entered in my objectives.

\section{Theoretical Physics in Arcetri}

L. B. \hspace{0.2 cm}
Which were the circumstances under which you really discovered your passion for theoretical physics?
\vskip 0.3 cm

L. M.  \hspace{0.2 cm}
There was a summer --- I think it was the summer of '63 --- when I went to visit a good friend of mine  who was in vacation in Vipiteno, on the Alps,  Ernesto Hofman, a physicist who later went to IBM, and I brought with me the paper by Gell-Mann about $SU(3)$ and the algebra of currents \cite{GellMann1962}. This paper was for me a discovery, an absolute discovery. 

I was in vacation but I would stay all mornings in the hotel studying Gell Mann's paper, trying to connect the various things. He described the origin of the symmetry applications to particle physics, the earlier paper by Fermi and Yang \cite{FermiYang1949}, the Sakata model \cite{Sakata1956} and his ideas about the $SU(3)$ symmetry. Gell-Mann's paper laid down a whole program for future investigations. I simply wanted to go along that program. And there could not be in Italy a better place to start than the school that was gathering in Florence around Gatto.

Raoul Gatto, born in Catania in 1930, grew up as a physicist in Scuola Normale di Pisa. After graduation he went to Rome as an assistant of Bruno Ferretti. Quite soon (in 1956) he left for the United States, to become a staff member of the Lawrence Radiation Laboratory in Berkley. 

The group of Luis Alvarez was in these years in full production, discovering lot of new hadrons with the hydrogen bubble chamber and Gatto absorbed quickly the exciting atmosphere of the laboratory. He made several papers on the phenomenology of the weak interactions (Fermi's imprint on all Italian theoretical physics) in particular on the weak decays of hyperons, based on the data collected by the Alvarez group. 
On his coming back, Gatto brought to Italy the new ideas concerning symmetry and group theory applied to particle physics, that were flourishing at the time in the US, a fresh air in the Italian theoretical physics.

Back in Italy, Gatto became the director of the newly formed theory group at the Frascati laboratories, where he found, as a junior partner, Nicola Cabibbo, freshly graduated with Bruno Touschek and recruited in Frascati by Salvini. 

Gatto gave recently a sharp description of his program of the time: {\it a research based on innovative theoretical ideas, in touch with the research programs of the big, international  laboratories, with the participation of recently graduated investigators, several of which have later obtained scientifically prestigious positions} (August 2011).\footnote{\dots ha cercato di mettere in atto una linea di ricerca basata su idee teoriche innovative pur restando connessa ai programmi dei grandi laboratori internazionali, facendovi partecipare giovani laureati, molti dei quali per il loro valore sono arrivati a posizioni scientificamente prestigiose (R. Gatto, unpublished. Courtesy of R. Casalbuoni).}
Nobody like Nicola could have filled this description any better.

\begin{figure}[ht]
\begin{center}
\includegraphics[scale=.50]{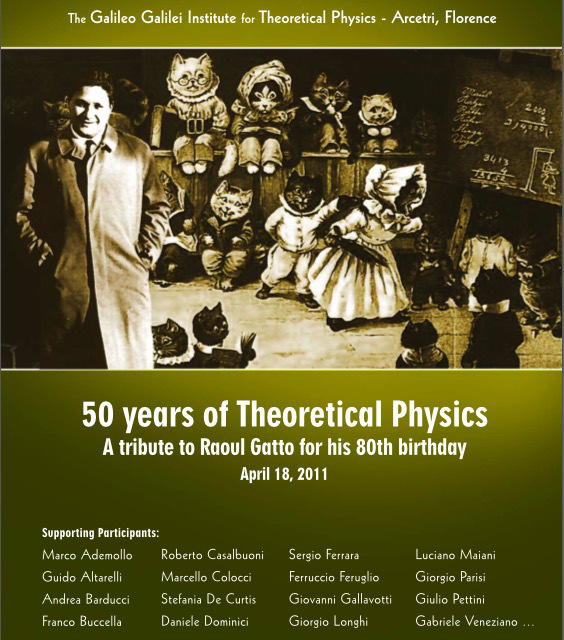}
\caption{{\footnotesize Poster to celebrate the eighty years of Raoul Gatto by his  pupils in Florence, Padova and Rome (see names on the poster). After many negotiations, unfortunately, the  Conference could not take place. The poster was created by Roberto Casalbuoni on the basis of a picture provided by Gabriele Veneziano. Courtesy of Roberto Casalbuoni.}}
\label{gattini}
\end{center}
\end{figure}

Gatto had a very interesting and attractive personality.  I went to Florence, met Gatto and said: ``I have not done anything with you. I have done  some experimental physics, but I want to move to theoretical physics. If you take me, I may come with my own fellowship because Ageno told me that I can continue the fellowship in Florence.'' Gatto simply agreed to accept me in his group, to have a desk, and for sometime we had no other direct relation (later, I learned from him that he had taken some information about me, consulting my thesis advisor, Cortellessa, before deciding).

Ageno had agreed to keep my fellowship at Sanit\`a even while I was working in Florence. I had simply to promise him that I would not lose contact. But that was easy, I agreed that I would come to Rome roughly every other week-end and keep him informed about what I was doing. Which in fact I did regularly all the time I was in Florence. 

I will always be grateful to Ageno for this decision and in fact I consider him next to my father, for what  he did for me. 

I moved to Florence at the beginning of March '64, just after graduating in Rome. The Romans there made a very interesting community. There was me, Buccella and Gallavotti who were the ``poor guys''. We were in a very simple boarding house which I had found through friends of mine  in Florence, friends from the summer vacations. The boarding house was just next to Santa Croce. It was called ``Pensione Pepi'',  it was in the old palace of the noble florentine family Pepi and was run by the marquise Pepi in person. The marquise was a very nice, very kind, old woman who, far from the brilliant family stories of the past, had to rent the rooms of a large apartment in the upper floor. 
Guido Altarelli, instead, had a wealthy uncle who had added for him an extra bonus to the fellowship. So, Guido could afford living in a more wealthy boarding house, ``Pensione Crocini'', closer to the centre. 

But I was very happy in Pensione Pepi. There were other people, besides us physicists, there were young student girls with whom we would talk during  the dinners we took all together. It was a very lively environment and those years have been very, very exciting. We were discussing all the time about mathematics, physics, theoretical physics and it was really wonderful. 

At that time the mathematician Beppe Da Prato, who had started as a physicist in Frascati with Cabibbo and with Gian de Franceschi, had moved from Rome to become a professor in Scuola Normale di Pisa. He often visited us in Florence. His strong personality influenced us a lot, mathematics-wise  (he introduced us to the mathematics of Hilbert spaces) and also politics-wise (a convinced communist, I was amused by Beppe's stories about Lenin loving to skate) but politics was anyway very far from my interests of that time.

Other people working with Gatto in Florence were: Marco Ademollo, Claudio Chiuderi, Enrico Giusti, Enrico Celeghini, Emilio Borchi, Mario Poli.  Gabriele Veneziano was writing his dissertation under the supervision of Gatto.

In Florence, Guido Altarelli emerged for his authority, clarity and sense of humour, and also for his  capacity to work with the Florence people: he worked with Longhi, became good friend of Ademollo and, in particular, of Chiuderi.

Gatto was masterly leading the large group of ambitious, young {\it gattini} as well as the somewhat older people he had found in Florence, Fig.~\ref{gattini}. 
 
 He was putting everybody in front of advanced but accessible problems (radiative corrections, $SU(3)$, $SU(6)$, $U(12)$, quark statistics, CP violation, weak interactions \dots you name it) he would discuss your results, send you back if not convinced, or write a draft paper.
 
We learned that we could compete with other groups, in US and Israel. Sid Meshkov defined us the {\it Italian mafia}, opposed to the {\it Israeli mafia} of Haim Harari and colleagues, who were working on the same subjects.

In everyday life, Gatto acted as the boss. He would come to your office and say: ``Oh, there is this calculation\dots \ it could be done, if you like.'' ``Yes, yes, yes, I will do it!'' And so I still remember when he assigned me my first calculation, an application of Gell-Mann's symmetry $SU(3)$ to neutrino reactions. I was very excited:  ``Yes, of course I will do it!''. This resulted in my first theoretical paper \cite{Maiani1964}.

\vskip 0.3 cm

L. B. \hspace{0.2 cm}
At that time, symmetry was becoming a central topic, but group theory had not yet really become a standard tool of theoretical physicists and I think that it was not taught as such. Where did your knowledge come from? 
\vskip 0.3 cm

L. M.  \hspace{0.2 cm}
In the first year, 1964, I was in Florence from February to summer. In between, Gian De Franceschi and I made a few seminars on group theory at Istituto Superiore di Sanit\`a. Preparing these seminars was the most efficient way to systematize what I had read here and there and make to myself a picture of what was the math involved and its applications to physics. The seminars appeared as an internal report of  Istituto Superiore di Sanit\`a (G. De Franceschi and L. Maiani, \textit{Introduction To Group Theory And Unitary Symmetry: A Course Of Lectures}, April 1964) and the editor of \textit{Fortschritte der Physik} proposed us to transform the notes  in a review paper.  So I stayed in Rome from September until Christmas, and we wrote a long article \cite{DeFranceschiMaiani1965}.  Of course, most of the mathematics was due to De Franceschi, he was very good at that. But the physics part was mine, meanwhile I had become proficient in the symmetry applications, Cabibbo theory and all that. It was a very interesting collaboration and the beginning of a long friendship. 

After Christmas '64 I went back to Florence and there arrived Giuliano Preparata. It was the beginning of a long collaboration. We were staying in the same boarding house for one year and later we rented an apartment to share, near Porta Romana. We started working together from the beginning and it was a very beautiful and very interesting bohemien period. 

In '64, in Florence, I had been a little isolated, because Altarelli, Buccella and Gallavotti had already worked with Gatto, so it was easier for them to continue. I remained a little aside, first of all because I had to keep up with theoretical physics. Gatto gave me to study  Bogoliubov's  \textit{Introduction to the Theory of Quantized Fields} \cite{BogoliubovShirkov1959} which I did, up to a certain point. I remember I studied up to Compton scattering, which closes the first part of the book, where Feynman diagrams are described. At that time, Gallavotti said: ``Look, after you have studied Compton scattering, you know everything you need. You don't have to study all the rest.'' There was an enormous part on renormalization which in fact I did not study until much later. I trusted what Gallavotti told me. 

I was isolated, and when Preparata arrived he was isolated too, so we joined forces. I liked to work with Giuliano, he was communicative like me, and very very good. We started a real friendship and we went into a closer collaboration with Gatto. 

In July 1964, Feza Gursey and Luigi Radicati \cite{GurseyRadicati1964},  had introduced a novel symmetry of the subnuclear particles extending the Gell-Mann and Ne'eman $SU(3)$ symmetry to $SU(6)$. 

The idea was to combine the spin symmetry of non-relativistic quarks, $SU(2)_{spin}$, with $SU(3)$ symmetry, postulating that bound states of quarks and antiquarks would be symmetric under the larger group $SU(6)\supset  SU(2)_{spin}\otimes SU(3)$.\footnote{This was reminiscent of the $SU(4)$ symmetry introduced by Wigner for nuclei, where the spin symmetry was combined with the isotopic spin symmetry, $SU(2)$, illustrated in {\bf Box 1}: $SU(4)\supset SU(2)_{spin}\otimes SU(2)$.}
 Baryons and baryon resonances did fit neatly  in a single, $56$-dimensional multiplet of $SU(6)$ and pseudoscalar and vector mesons in a single $35\oplus 1$-dimensional $SU(6)$ complex. 

The Florence School under the guidance of Gatto attacked with enthusiasm the exploration of the newly discovered  $SU(6)$ symmetry and its relativistic extensions.

\begin{figure}[ht]
\begin{center}
\includegraphics[scale=.65]{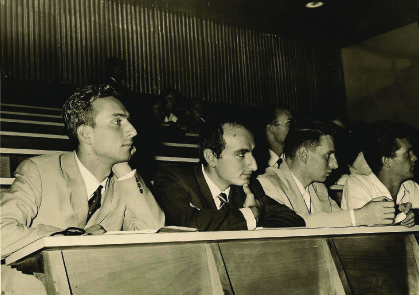}
\caption{\footnotesize{From left: Luciano Maiani, Antonio Degasperis, Sergio Doplicher and Gian De Franceschi, attending a Conference in Pisa, 1964.}}
\label{pisa}
\end{center}
\end{figure}

\section{Baryon Resonances with Gatto and Preparata}

In summer '65, Giuliano and I went for the first time to the US, to attend the Brandeis Summer School, near Boston, with many other Italians, including Buccella De Franceschi and other Roman colleagues.

It was a great experience. There I made friendship with Samuel Ting, who was working at Columbia University and, above all, I met Nicola Cabibbo. He was one of the lecturers, the superstar of the moment for his theory of the universality of weak interactions (Box {\bf 1}), so we were actually looking at him as a semi-God. He was interested to discovery and a very nice person. Giuliano and me enjoyed several conversations with him and with Nicola's wife Paola, who was then expecting their son. 

In fact, in Brandeis, it was Giuliano who introduced me to Nicola and Paola. Giuliano knew Nicola independently, because Nicola had been with Giuliano's brother at the university. Nicola knew from his friend that he had a younger brother who was becoming a brilliant theoretical physicist.  

 From the Brandeis courses we brought back to Florence many new ideas. Those, in particular  that had come out from the lectures of Benjamin W. Lee. 
 
 Ben Lee spoke about  saturation of the current algebra relations with the lowest lying baryons, octet and decuplet, which gave the well known $SU(6)$ prediction for the weak coupling $g_A/g_V=5/3$, a value substantially larger than the experimental value $g_A/g_V\sim 1.25$. We said:  ``Why don't we try to add the next resonances and see if they improve the situation?''.   We proposed the idea to Gatto, who found it interesting and encouraged us  to go on. And so Giuliano and I got engaged in this very complicated calculation.

\vskip 0.6 cm 
\begin{tcolorbox}[breakable, enhanced]

\footnotesize{
{\bf Box 1. Isospin, Weak Isospin and the Cabibbo angle}\\
\vskip 0.1cm
In 1932, Werner Heisenberg, motivated by the near equality of proton'a and neutron's masses, introduced the concept of {\it Nucleon}, a quantum system made of two states, proton and neutron, similar to the electron which can exist in two spin states, spin $up$ and spin $down$ \cite{Heisenberg1932}. In 1937, Eugene Wigner  \cite{Wigner1937} introduced transformations which mix Nucleon states into one another:  
\begin{eqnarray}
&&N=\left(\begin{array}{c} p \\ n \end{array}\right) \to U~\left(\begin{array}{c} p \\ n \end{array}\right)= UN~({\rm isospin~transformation})\label{isospin} \nonumber
\end{eqnarray}
$U$ is an arbitrary, complex, $2\times 2$ matrix with unit determinant. Going further, Wigner made the hypothesis that nuclear interactions are left invariant by these transformations and coined the nam {\it isotopic spin symmetry}, 
which would then be a symmetry of the nuclear levels, much the same as rotations are a symmetry of the atomic levels.


Fermi used the concept of  Nucleon in his theory of Weak Interactions \cite{Fermi1934a} \cite{Fermi1934b}. The latter act on the Nucleon, the ``heavy'' particle (baryon), ``raising'' the neutron into a proton. At the same time a pair of ``light'' particles (leptons) is created, which can be seen as ``lowering'' a negative energy neutrino into a positive energy electron. In total: $n\to p ~e^- \bar \nu_e$.

To include the observed violation of parity, Feynman, Gell-Mann  and others reformulated the Fermi theory assuming that only left-handed particles participate in the weak interactions, thereby introducing the {\it weak isospin} symmetry acting on doublets of left-handed fields only:
\begin{eqnarray}
&& N_L=\left(\begin{array}{c} p \\ n \end{array}\right)_L;~({\ell}_e)_L
=\left(\begin{array}{c} \nu_e \\ e \end{array}\right)_L\nonumber\\
&& N_L \to U_W N_L;~({\ell}_e)_L \to U_W ({\ell}_e)_L\nonumber
\end{eqnarray}

Including the left-handed muon doublet 
\begin{equation}
({\ell}_\mu)_L=\left(\begin{array}{c} \nu_\mu \\ \mu \end{array}\right)_L;~({\ell}_\mu)_L \to U_W ({\ell}_\mu)_L~({\rm same}~U_W)\nonumber
\end{equation} 
Feynman and Gell-Mann used the isospin symmetry of Heisenberg to explain the surprisingly near equality of the Fermi constants of neutron and muon decays \cite{FeynmanGellMann1958}. Harmony was broken by the weak decays of strange particles, e.g. $\Lambda \to p ~e^- \bar \nu_e$: a decay similar to neutron decay, but with a Fermi constant about $0.22$ times the neutron one. 

How could one reconcile the lack of universality of $\Lambda$ decay with the universal neutron and muon decays? The answer came from Nicola Cabibbo \cite{Cabibbo1963}.
The argument is best framed in the language of quarks \cite{Gell-Mann1964}. 

Extending the strong interaction symmetry from isospin to the ``Unitary Symmetry'' of Gell-Mann and Ne'eman, $SU(2)\to SU(3)$ and one brings in, as the basic constituents of hadrons, {\it three quarks} with spin $1/2$ and fractional electric charges, $Q$
\begin{eqnarray}
&& q=\left(\begin{array}{c} u \\d \\s  \end{array}\right),~Q=\left(\begin{array}{c} +2/3 \\ -1/3\\-1/3  \end{array}\right) \label{quark3} \nonumber
\end{eqnarray} 
The {\it strange} quark, $s$, carries zero isospin and the negative unit of strangeness.  
Hadron weak decays are due to the weak quark decays (recall that: $n=[udd],~p=[uud],~\Lambda=[uds]$):
\newpage

\begin{itemize}[leftmargin=.3in]
  \item [] $d\to u~e^- \bar \nu_e$, which induces neutron's decay, $n\to p$ transition;
  \item [] $s\to u~e^- \bar \nu_e$, which is responsible for strange hadron decays, e.g. the $\Lambda \to p$ transition.
\end{itemize}

Cabibbo observes that the quark fields reported above have definite mass, defined by the strong interaction. However, weak interactions need not respect the strong interaction classification. Since $d$ and $s$ quarks have the same electric charge, a quantum number respected by both interactions, the quark coupled to the $u$ quark in weak decays could be  a {\it superposition} of the two. The weak quark doublet would be:
\begin{eqnarray}
&&(q)_L=\left(\begin{array}{c} u \\ d_C \end{array}\right)_L,~ (s_C)_L
\nonumber \\
&& d_C=\cos\theta d+\sin\theta s,~s_C=-\sin\theta d + \cos\theta s \nonumber 
\end{eqnarray}
$\theta$ is a universal constant, the {\it Cabibbo angle}, and $(s_C)_L$ is a weak isospin singlet, not participating in the weak interactions, similar to the right-handed fields. In conclusion, Cabibbo's complete weak isospin scheme of quark and leptons is (assuming no right-handed neutrinos):
\begin{eqnarray}
&& \left(\begin{array}{c} u \\ d_C \end{array}\right)_L,(s_C)_L,u_R,d_R,s_R;~
\left(\begin{array}{c} \nu_e \\ e \end{array}\right)_L,e_R;~\left(\begin{array}{c} \nu_\mu \\ \mu \end{array}\right)_L,\mu_R.
\nonumber
\end{eqnarray}

Weak decays of strange baryons and mesons are very well reproduced by $\sin\theta = 0.225$.

 }
 \end{tcolorbox} 
\vskip 0.3 cm


 We worked quite a while, trying to solve a complicated system of equations, and then all of a sudden a very simple result came out, which also agreed with the experimental value of $g_A/g_V$! It was an absolute miracle. We went to discuss with Gatto and he was very excited. The discussion resulted in two joint publications \cite{GattoMaianiPreparata1966,GattoMaianiPreparata1967}.
  
 The result was interesting because connecting the lightest baryons with the higher resonances gave a result of a new type, which went beyond the exact $SU(3)$ or $SU(6)$ relation. So we were very excited. Even more, when we learned that Nicola Cabibbo, who was in Geneva, had been able to reproduce our result in an extremely simple way \cite{CabibboRuegg1966}. With Henri Ruegg, in a 2 pages long article, he had given a synthetic group theoretical argument which led to our result. 
 
 So the solution of our complicated calculations could be described in few lines of group theory! It was another miracle and it gave a better perspective to our solution. I was really impressed by Nicola's insight in physics and skill in group theory.

At that time, Nicola was thinking of coming back to Rome, after the US and CERN.  And  this opened up new perspectives. 

In Florence, we continued working on the symmetries of elementary particles. At that time there was the triumph of the so-called $U(12)$ symmetry proposed by Abdus Salam  \cite{SalamEtAl1965}, then directing in Trieste the Centre for Theoretical Physics (ICTP) he had just founded.

 In 1965, several attempts had been made towards the problem of extending to the relativistic domain the $SU(6)$ symmetry of Gursey and Radicati. 
 
That of Salam was a  bold attempt to make $SU(6)$ symmetry into a relativistic theory. After this attempt everybody said: ``We have to work on that!'' Then we, too, started to work on that. We were very excited and Giuliano and I with Gatto, in summer 1966 in Erice, even made a weird attempt in which we went up to $SU(48)$ or something like that.

Unfortunately, shortly after, Sydney Coleman put an end to all the excitement. With Jeffrey Mandula, he proved a famous \textit{no-go theorem} that said that any theory that combines relativity with an internal symmetry cannot contain interactions \cite{ColemanMandula1967}. 

Interactions between particles are the very essence of physics, the reason why the Sun shines, the sky is blue and we see the world the way we see it! Relativistic $SU(6)$ was simply a blind alley and Coleman's result was the end of the story. 
\begin{figure}[t]
\begin{center}
\includegraphics[scale=.85]{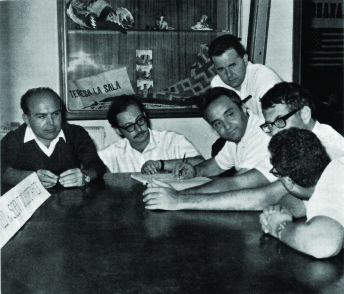}       
\caption{\label{fig:nicetal}  {\footnotesize   Making the Standard Theory, Erice School in Subnuclear Physics, 1967. Going around the table from left: Bruno Zumino, Sidney Coleman, Nino Zichichi, Nicola Cabibbo, Sheldon Glashow and Murray Gell-Mann \cite[p. 148]{courier}.}}
\end{center}
\end{figure}

With the end of 1966, the Florence experience came to an end. Although Gatto had done very well in Florence and had became really famous for having grown all these pupils, he was thinking to move.  In 1967 he went one year in Geneva and later on, in 1968, he would move to Padova.  I got a permanent position in the Istituto Superiore di Sanit\`a and Ageno made me understand that it was time for me to come back.

Other Romans started to move, in particular to the States, where Altarelli went in 1967. The perfect atmosphere of  '65/'66 was not there any more.

Giuliano and I left the apartment near Porta Romana and moved in a new boarding house which fortunately was at a higher level than Santa Croce, so we were left undamaged by the Florence big flood, which arrived on November 4th. 

It was time to change and, towards the end of the semester, I went back to Rome.   

What did we achieve in Florence? Certainly we participated in the struggle of theoretical physics of those years towards a theory of the strong interactions and we had been recognized as useful interlocutors. 

The main theme of our research was to go beyond the pure concept of symmetry, $SU(3)$, $SU(6)$ or higher, and  delve deeper into the role of quarks. It was in these years that the problem of quarks and statistics emerged. 

The lowest lowing baryons are bound states of three quarks in $S$-wave. However, the three quarks are in a {\it symmetric} configuration with respect to spin and flavour quantum numbers, the latter being the three quark labels, $u, d, s$ encoded in the labels of the $SU(3)$ representation. The problem is seen most clearly with the state called  $\Delta^{++}$, the resonance first observed by Fermi in the early 1950s. In terms of quarks, one has $\Delta^{++}=u^\uparrow u^\uparrow u^\uparrow$, the arrow indicating spin up. It is obvious that, in the absence of further labels, this is a symmetric state under the exchange of any two quarks, which is incompatible with quarks having spin 1/2 and therefore having to obey the Fermi statistics\footnote{The difficulty revived ideas about the so-called {\it parastatistics} of order $N$, in which $N$ identical particles are allowed to fill one quantum state. In Fermi statistics $N=1$, O.~W.~Greenberg~\cite{Greenberg1964} proposed quarks to obey a parastatistics of order $N=3$.\footnote{For a general analysis see \cite{GreenbergerMessiah1965}.}
 The idea was later realized in QCD, with quarks being given an $SU(3)_{color}$ label e.g. $u\to u^\alpha,~\alpha=1,2,3$. In this case, the color singlet state of three $u$ quarks: $\epsilon_{\alpha \beta\gamma} u^\alpha u^\beta u^\gamma$, with $\epsilon_{\alpha \beta\gamma}$ the Levi-Civita tensor, is antisymmetric under exchange of two $u$ fields. So, three $u$ quarks can occupy the same spin and space wave function without violating Fermi statistics, similar to the two electrons that can be in the same orbital, due the two-valued spin quantum number. First ideas about parastatistics go back to~\cite{Gentile1940}.}.

Gatto, Giuliano and myself had hit precisely this problem with our attempt to introduce a mixing between proton and $\Delta$ with higher resonances. We choose a $20$-dimensional representation, which had the lowest dimensionality and, in addition, was antisymmetric as appropriate for fermions. However, to be consistent with the lowest lying baryons, one had better chosen the $70$ dimensional representation --- as our competitor Haim Harari did \cite{Harari1966a} \cite{Harari1966b} --- which has a mixed symmetry and, with one unit of internal orbital angular momentum, would reproduce a symmetric configuration. 

With the $70$ dimensional representation there were too many parameters and one would not obtain any particular relation for the weak interaction constants of the baryons. But it was the rational choice. 

Anyway, our work did put attention on the chiral group $SU(3)\otimes  SU(3)$, and showed that saturation of the commutation relations with higher resonances could correct the predictions of $SU(6)$, which was lacking theoretical support after the Coleman-Mandula theorem.

In these years, many people interpreted the quark statistics problem as an indication that quarks were a purely mathematical device for the bookkeeping of the hadrons  internal quantum numbers. 

The puzzle remained until the end of the 1960s and the first 1970s, when it was clarified by the observation of deep inelastic scattering reactions, which showed the reality of spin $1/2$ quarks, and required the introduction of a further quark quantum number, {\it color}. Color restores the overall antisymmetry under quark interchange and reveals itself as the true basis of the fundamental strong interactions of quarks, carried by particles similar to the photon, the gluons, hitherto unsuspected. 

Coming back to the results obtained by the Florence school, the electrodynamic calculations concerning electron-positron annihilation were there to stay and to be used by the experimental collaborations, in Frascati and elsewhere. 

By far the most important result was, however, the so-called {\it Gatto-Ademollo theorem}, the statement that deviations from exact $SU(3)$ symmetry intervene in baryon and meson weak decays only to second order in the symmetry breaking parameter \cite{AdemolloGatto1964}. This result had an important role in the analysis of the experimental results and is today widely accepted to justify the excellent agreement of the Cabibbo theory with baryon and meson weak decays.

\vskip 0.3 cm 

L. B. \hspace{0.2 cm}
In early June 1967 you and Preparata submitted your first article written in collaboration with Nicola Cabibbo \cite{CabibboMaianiPreparata1967}. It was on the radiative corrections to $\pi-\beta$-decay\dots

\vskip 0.3 cm

L. M.  \hspace{0.2 cm}
Yes, meanwhile Nicola had in fact come back to Italy from Geneva. In 1967 he got a chair of Theoretical Physics in L'Aquila and  was called in Rome the following year.  And so, quite naturally, Giuliano and I started to work with Nicola and we wrote what I still consider a very interesting paper. We observed that with fractionally charged quarks the radiative corrections to the Fermi theory were infinite. However, one could get a finite result if quarks had integer charges. We went even further, proposing a model in which one had 3 multiplets of quarks with different integer charges in such a way that the average was resulting in a fractional charge. I was very excited and Nicola very happy. It was brilliant indeed, but soon we learned  that it had already been done! We had  rediscovered the {\it Han-Nambu model} \cite{HanNambu1965}  that we had called the SUB-model. Of course the result on the radiative correction was original and it attracted a lot of attention and I made a talk about it at the International Particle Conference in Heidelberg. 

Heidelberg was my first conference talk. Nicola was also there. One afternoon we left the Conference and went out with his car driving along the Rhine valley, tasting {\it Schnaps} here and there, talking about everything\dots \ I was really fascinated by his personality. That little travel established a friendship that lasted many decades. Meanwhile Preparata had got married and went to the States with a Fulbright fellowship. 

In autumn '67 Giuliano settled in Princeton, where he made a very good, very interesting paper with William Weisberger \cite{PreparataWeisberger1968}, which was in a way a continuation of our paper. They demonstrated what is now called the non-renormalisation theorem of Preparata and Weinsberger. It was a very important result which applied to the weak-interactions, provided strong interactions were mediated by an electrically neutral vector meson, that had been called ``the gluon''.  


Preparata and Weisberger paper was where I learnt about the gluon which, at that time, was a highly hypothetical particle, to mediate highly hypothetical strong interactions between quarks. They showed that the results of the simple quark model, which we had used in our work with Nicola,  would remain valid in the presence of strong interactions of this particular form. At the time, nobody could have guessed that only few years later the world recognised strong interaction would have been precisely of this form, only with eight, rather than one, neutral gluons. This is QCD, in which the Preparata-Weisberger (PW) theorem could be extended without changing a word. 

Ironically, the PW theorem did not apply to the Han Nambu theory, since there the gluons are charged, but in QCD it supported the GIM mechanism proposed by Glashow, Iliopoulos and myself, as we will see later.

To close on the radiative corrections to weak decays, another paper appeared at about the same time, by Alberto Sirlin \cite{Sirlin1968}. It showed that the divergence present with fractionally charged quarks could be eliminated if the Fermi interaction was the low energy limit of an interaction mediated by a charged intermediate vector boson. So, as I noted in my talk in Heidelberg, by requiring finite radiative corrections to $\beta$ decays, one was confronted with two alternatives. Either Fermi's  was really a local interaction and one should  have integrally charged quarks, as in the Han-Nambu or SUB model, or one had fractionally charged quarks and then the intermediate vector boson had to exist. We know now, after the electron and neutrino deep inelastic scattering data and the discovery of the $W$ and $Z$ bosons, that Nature goes along the second alternative.
\vskip 0.3 cm

\section{Difficulties with the Weak Interaction Theory}

L. B.  \hspace{0.2 cm}
And so we are now in '68 and you had started to work systematically with Nicola Cabibbo\dots
\vskip 0.3 cm

L. M.  \hspace{0.2 cm}
We did works on  symmetry  breaking\dots\  At that time the glamour were the papers by Gell-Mann, Oakes and Renner \cite{GellMannOakesRenner1968} and by Glashow and Weinberg \cite{Glashow 1968} who showed that the $SU(3)$ symmetry is spontaneously broken, generalizing to $SU(3)\otimes SU(3)$ the old idea of Nambu and Jona Lasinio. It was the beginning of the Standard Theory. The vacuum is not invariant\dots \ I had to learn about the Goldstone theorem and so on. So it was really very exciting. 

I spent the summer '68 in Boulder, Colorado, at the Boulder Summer School, finishing a review paper written in collaboration with Giuliano about the algebra of currents \cite{MaianiPreparata1969}. In the same school there were Carlo Bernardini and his wife Silvia so I had the opportunity of knowing them a little better. Then I went to SLAC, Stanford, to visit Giuliano Preparata who had moved there from Princeton. Finally, on may way back to Italy, I stopped a few days in Washington, where I had an old friend of mine, Sydney Meshkov, who was at the National Bureau of Standards (NBS), now called the National Institute of Standards and Technology (NIST). There was a physics department in NBS and it was quite similar to the physics laboratory of the Istituto Superiore di Sanit\`a, except on a bigger scale. 

I spent just a few days in Washington, in the second part of August '68, discussing about physics and physicists with Sidney. At one moment, he gave me a paper, saying: ``Hey, there is this new paper by Gell-Mann, Goldberger, Kroll and Low \cite{GellMannEtAl1969}. Referring to the authors, he used the expression ``an impressive firepower''. 
  
The story was that few months before there had been a paper by Ioffe and Shabalin \cite{IoffeShabalin1967} and also one by Francis Low \cite{Low1968} that showed that if you try to compute the higher-order corrections in the weak interaction with the vector-boson theory, you discover that these higher corrections would produce decays like  $K_L\to \mu^+ \mu^-$ or the mixing between the two neutral Kaons, $K^0$ and its antiparticle $\bar K^0$, to produce the two observed neutral mesons called $K$-short and $K$-long. These processes are known as ``flavour changing neutral current'' effects (FCNC). In the weak interaction theory, FCNC processes would be mediated by a neutral vector boson coupled to an electrically neutral current.  Experimentally, FCNC processes were known to be very much suppressed, or even, at the time, compatible with vanishing (see  {\bf Box 2}). 

 Joffe-Shabalin and Low's papers showed that  to reproduce this situation, even in a theory with only one charged Intermediate Boson, you had to introduce a very small ultraviolet cut-off, of  about 3 GeV, which was totally incomprehensible.  
 
 The meaning of the cutoff is to give an energy scale above which the theory looses its validity and has to be drastically corrected. 
  However, energies of 3 GeV had been amply reached at the time by particle accelerators and nothing strange had been seen to happen there, in particular in neutrino reactions which were produced by the same weak interactions considered in these papers\footnote{An extensive account on the evolution of neutrino experiments, from the 1960s to the discovery of neutral currents in 1973, has been given by Dieter Haidt in~\cite{Haidt2004} and in~\cite[pp. 41-54]{Cashmore 2004}.}. 
  
This was the mystery\dots \ But Sidney gave me also the paper which said that somebody had found a solution. However, the solution was remarkably complicated, by having cancellations between a huge number of vector bosons. It looked very artificial. Anyway, I got the problem. 

The last day that I was in Washington I went to see the White House. I was just walking in front of the White House when it started raining, but I said: ``I don't care to get wet.'' I was happy to walk in the rain, because I had an interesting problem to think about and also because, after a long summer alone, I was going back to Italy where I had a sweetheart, Pucci de Stefano, whom I really loved. In fact one year later she became my wife. So I could be very happy indeed! 
 
I came back to Rome and, after one sleepless night, took my car and drove to Forte dei Marmi, a well known seaside resort in Tuscany. Pucci was there with her family and she was very surprised to see me appearing at her door at 7 o'clock in the morning. So, we had vacation together and it was there that I asked her to marry me.  

Back in Rome, in September, I went to the International Conference of High Energy Physics (ICHEP) in Vienna. It was a very important conference, where the deep inelastic scattering was presented, with the observation of Bjorken's scaling, the beginning of the discovery of quark reality. 

There I found Nicola acting as convenor of the weak interactions. He was very excited by the paper by Ioffe and Shabalin  and by the new result of Gatto and collaborators that had shown that another divergent higher order amplitude, which required an even smaller cutoff, could be cancelled by a special value of the Cabibbo angle \cite{GattoSartoriTonin1968}. 
  
  This was really a fantastic result and we started working on that. 
\vskip 0.3 cm
\begin{tcolorbox}[breakable, enhanced]
\footnotesize{
{\bf Box 2. Fermi Theory in Higher Orders}\\
\vskip 0.1 cm
Fermi described the weak interactions as a contact interaction between four fermion fields (e.g. $G_F[\psi_n\psi_p^\dagger\psi_\nu \psi_e^\dagger]$) with a coupling constant $G_F$, the Fermi constant, with dimension of [mass]$^{-2}$. As such, the interaction was known to be {\it non-renormalisable}. Common feeling was that Fermi's had to be the low energy limit of a more complete, {\it renormalisable}, theory which should enjoy the same status of quantum electrodynamics (QED), whose consequences can be computed in perturbation theory in terms of few physical parameters (masses and coupling constants). 

An early step was to assume a charged intermediate vector boson, analogous to the photon, to mediate Fermi interactions. The idea evolved into the massive Yang-Mills theory with weak and electromagnetic unification,  proposed in 1961 by Glashow \cite{Glashow1961}. This line evolved further  with the work of Brout and Englert  \cite{EnglertBrout1964} and of Higgs \cite{Higgs1964a,Higgs1964b}, who identified the mechanism of spontaneous symmetry breaking to give a mass to the vector bosons of the weak interactions. Finally, Weinberg \cite{Weinberg1967} and Salam \cite{Salam1968}  incorporated spontaneous symmetry breaking in a realistic model of the electromagnetic and weak interactions. Demonstration of renormalisability of the Weinberg-Salam model is due to 't Hooft and Veltman \cite{tHooftVeltman1972}. Similar to Glashow's, the model of Weinberg and Salam was able to describe only the leptonic sector of the weak interactions. This was what could be called the {\it top-down} approach.

There was however also a {\it bottom-up} approach, initiated at the end of the 1960s, that consisted in exploring the higher orders of the Fermi theory, extended by Cabibbo to strange particle decays and strongly supported by the experimental data (for an illuminating reconstruction of the bottom-up approach, see \cite{Iliopoulos2016}).

Assume to compute the weak corrections to some amplitude, to any order in $G_F$, using the Fermi interaction. The theory being non-renormalisable, as you go to higher order in $G_F$ you will encounter divergences of higher and higher order in the integration over the momenta of internal particles. Assume then that you cut the integration over virtual momenta at some fixed, finite momentum $\Lambda$. Now there are no infinities and the perturbation series can be rearranged in powers of the cutoff $\Lambda$ in a way dictated by the physical dimension, [mass]$^{-2}$, of the Fermi constant, to give something like:

\begin{eqnarray}
&& A=\sum_{n=0}^\infty~(G_F\Lambda^2)^n~A^{(0)}_n+(G_F M^2)~\sum_{n=0}^\infty~(G_F\Lambda^2)^n~A^{(1)}_n+\cdots \nonumber \\
&& +\cdots ~(G_F M^2)^k~\sum_{n=0}^\infty~(G_F \Lambda^2)^n~A^{(k)}_n+\cdots  \nonumber\\
&&=(G_F\Lambda^2)~A^{(0)}_1+\cdots +(G_F M^2)~A^{(1)}_0 +(G_F M^2)(G_F \Lambda^2)~A^{(1)}_1 +\cdots \nonumber 
\end{eqnarray}

$A$ is the desired amplitude, $A_n^{(k)}$ calculable quantities of order unity and M$^2$ is a dimensional normalising parameter, which we fix to be $M=1$~GeV (with this choice, $G_FM^2\sim 10^{-5}$). 

The value of $\Lambda$ corresponds to the energy scale where the Fermi theory has to be changed into the ``right'' high energy theory. This should happen where the series in powers of $G_F\Lambda^2$ will start diverging, that is when $G_F\Lambda^2=1$, corresponding to $\Lambda \sim 300$~GeV. The same applies, approximately, if we replace the Fermi interaction with the more convergent, but equally non-renormalisable, Intermediate Vector Boson theory. 

In the last part of previous formula, we have reported the leading terms of each category, which illustrate what can go wrong with the $G_F\Lambda^2=1$ philosophy. {\bf The scene is set by the second term}, which is the lowest order weak interaction amplitude. The structure of the weak amplitudes was well tested, in the 1960s, by the success of the $V-A$ theory, for lepton decays, and the Cabibbo theory for hadron decays. Weak amplitudes were known to violate several strong interaction symmetries, like Parity (P), Charge Conjugation (C) and isotopic spin, and to violate strangeness conservation by one unit in semileptonic, charged current processes, the so-called $|\Delta S|= 1$ rule. The latter selection rule was tested  to high precision by stringent upper limits to the rate of the decay $K_L \to \mu^+ \mu^-$ (strangeness violation in a neutral current process) and by the extreme smallness of the $K_L-K_S$ mass difference (a $|\Delta S|=2$ amplitude). Both processes would occur only to second order in Cabibbo's theory.

The {\bf first term} represents a weak correction to the generic strong interaction amplitude, $A^{(0)}_0$. In the  $G_F\Lambda^2=1$ limit, it would introduce P, C, isospin and strangeness violations of {\cal O}(1) in the strong interactions. On the contrary, observation would require, at first sight, an absurdly small value of  $\Lambda$.

The {\bf third term} contains weak corrections which are formally of the second order.~As shown by Ioffe and Shabalin \cite{IoffeShabalin1967} and by Low \cite{Low1968}, to comply with the limits provided by $K_L \to \mu^+ \mu^-$ and by the  $K_L-K_S$ mass difference, a value of the cutoff as low as $\Lambda \sim 3$~GeV is required.
}
\end{tcolorbox}

\vskip 0.3 cm

We made a first paper \cite{CabibboMaiani1968}, where we thought we had improved on Gatto and had understood things better. 

We continued working on this subject during the whole 1969.
Part of the results found by us have been superseded by the following investigations, but one still remains, incorporated in the Standard Theory, so let me briefly illustrate what we did, also as an introduction to later developments.

 Like Gatto {\it et al.}, we did start from the most leading divergent term, that could endanger the very structure of the strong interactions, the first term in the last equation given in {\bf Box 2}. It had been already shown \cite{BouchiatIliopoulosPrentki1969} \cite{Iliopoulos1969} that this term affected only the part of the strong interactions, which breaks the chiral  symmetry $SU(3) \otimes SU(3)$. They had also shown that, if the breaking has the transformation properties of the quark mass terms, the dangerous terms can be expressed as four-divergences of vector and axial currents. Four-divergences can be notoriously eliminated from the Lagrangian and therefore the dangerous terms are in fact innocuous, even with the condition $G_F\Lambda^2=1$. 
 
 We went a bit further. We retrieved an old result \cite{FeinbergKabirWeinberg1959} \cite{CabibboGattoZemach1960}  stating that {\it any} combination of terms quadratic in the quark fields (i.e. generalised mass terms) can be diagonalised by left and right multiplication of two suitable unitary matrices. These transformations belong to chiral $SU(3) \otimes SU(3)$ and leave invariant the chiral symmetric part of the interaction. In other words: {\it no matter what arises from the leading weak corrections, provided it is of the form of a quark bilinear, a definition of P, C and strangeness can always be found such that these symmetries are conserved by strong interactions}.

 We noted, however, that the resulting diagonal quark masses had necessarily to  break isotopic spin. This was an interesting result: it had been always assumed that isospin was an exact symmetry of the strong interactions, broken only be electromagnetic exchange of photons. The divergent weak interactions produced a second source, which could give observable effects. In fact, precisely in these years, cases had been found indicating that there are isospin violating processes which {\bf are} unaffected by photon exchange, in the exact chiral symmetry limit: one is the $\eta \to 3 \pi$ decay amplitude \cite{Sutherland1966}  and the other is an isospin violating combination of $K$ and $\pi$ mass differences \cite{Dashen1969}. Our observation provided an independent source, not arising from photon exchange, that could explain these processes, at least qualitatively. 
 
 Using Dashen's sum rule for meson mass differences, we could determine the $up$ and $down$ quark masses separately. We found surprisingly small and surprisingly different masses: $m_u \sim 5$~MeV and $m_d\sim 7$~MeV, using $m_s\sim 150$~MeV. 
 
 It was thought  previously that approximate isospin symmetry was due to a small fractional difference between the masses of the $up$ and $down$ quarks, while we found about 100\% difference. The reason of approximate isospin lies rather in the fact that both $up$ and $down$ quark masses are very small on the hadronic scale, of order $1$~GeV, and isospin symmetry is recovered in the limit of vanishing $up$ and $down$ quark masses. 
 
 The $down$-$up$ positive mass difference, by the way,  neatly explained why the neutron is heavier than the proton, a fact of great cosmological significance, and something never reproduced by electromagnetic self energies, which tend to make the proton heavier.
 
 Our attention, however was concentrated on the value of the Cabibbo angle. 
 
 To eliminate parity and strangeness violations arising from the divergent weak corrections to quark masses, one had to make, as we have seen, a chiral rotation. In doing so one would change the angle of the weak interactions and also the values of the quark masses. We observed that there was a special value of the angle which made so that after including the weak and the electromagnetic corrections and diagonalising the result, one did come back to the initial values of the physical quark masses. The magic value of the angle, expressed in terms of the pion and kaon masses, was surprisingly close the real value. Our result was approximately the same as the one found by Gatto \textit{et al.} \cite{GattoSartoriTonin1968} and could be expressed as: 
 \begin{equation}
 \sin\theta \sim \sqrt{\frac{m_d}{m_s}}\sim 0.22 \nonumber
 \end{equation} 
 a very successful relation, indeed.
 
Our was a sort of self-consistency condition (equal initial and final quark masses), in line with the ``bootstratp'' ideas, promoted by Chew and Mandelstam at that time and much valued by Nicola in the context of the strong interactions. We published our results in detail on the \textit{Physical Review}  \cite{CabibboMaiani1970a} and in the book to celebrate Edoardo Amaldi's sixtieth birthday \cite{CabibboMaiani1970b}.

\vskip 0.2cm

At the end of 1968, I had made applications to various places in the United States, for a one year postdoc position. My plan was to get married in 1969 and move to the USA immediately after, as Giuliano had done two years before. Pucci was not very happy, but she finally understood that this was really an essential step for me and accepted the idea. 

I received two offers, one from Caltech and the other from Harvard. I remember very clearly being in the kitchen of my mother's apartment where I was living, eating lunch by myself with the letters in front of me and trying to decide.  I made this argument: ``Caltech is fantastic, but I will never be able to work with Feynman or Gell-Mann there; they are too busy''. At Harvard, there was Sydney Coleman, who had just spent one year in Rome, I knew him and  I thought that he was also fantastic. With Syd Coleman I could work, and there was also Sheldon Glashow, whom I had met at the Varenna School (in 1964) and who was working on the weak interactions (I learned only later that Glashow  had been the referee of the {\it Physical Review} paper with Nicola on the calculation of the Cabibbo's angle by cancellation of the divergences; his comment to the \textit{Physical Review}: ``English is terrible, but physics is good.''). 

So, this is why I accepted Harvard, without knowing that John Iliopoulos, who was also working on the same subject, had just accepted to go there. In fact that was an incredible stroke of luck. I was going precisely where there were the two people who had known all the story, had contributed to the story and it was possible there to simply continue on the road I had taken with Nicola. 

Not a secondary reason for accepting Harvard's offer was that Boston seemed to me a more `european' destination than Caltech, thinking that this would be a softer transition for Pucci, who had never been in the States before and was somewhat diffident towards the american way of life (with the Black Panthers movement and the Manson affair).

\section{GIM Mechanism}
\label{gimdisc}
 
 L. B. \hspace{0.2 cm}
What were your expectations now that you were on the verge of moving to Harvard?
 \vskip 0.3 cm

\noindent L. M.  \hspace{0.2 cm}
Pucci and I got married in July '69, and we arrived in Harvard in November. Going to Harvard was extremely exciting. Cambridge  is a very nice place,  and we had found a house which was in a way very Romantic, an apartment under the roof in a wooden house. It was in a reasonable part of Cambridge, not far from Harvard Square, one could even walk to the Department. At the corner of the street we had  a simple fish place, called the Legal Sea Food, which was the preferred fish place by our physicist friends. 

Now the Legal Sea Food has become very famous, with many elegant branches. At that time it was simply a fish shop where you could sit down and eat  the fish of your choice, simply cooked by them. We also loved to buy a pair of lobsters and cook them in our place. While we were in Harvard, we went there often with Shelly and John. 

Two places were frequented by the Harvard physicists, one was the Legal Sea Food and the other was a Chinese restaurant in Medford, the ``Peking on the Mystic'', with a delicious Peking duck.
\begin{figure}[ht]
\begin{center}
\includegraphics[scale=.85]{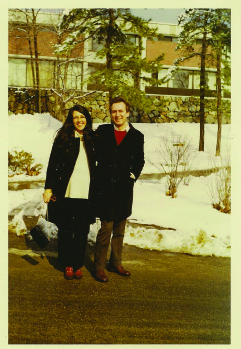}
\caption{\footnotesize{With Pucci in the campus of Brandeis University, winter 1969-1970.}}
\label{brandeis1969}
\end{center}
\end{figure}

When I arrived in Harvard I was still taken by my work with Cabibbo. Shelly and John had strong doubts about it, however, and we discussed at length about it. Basically, I think they thought that the $G_F\Lambda^2$ philosophy was just a tool to guess the correct low energy theory, but that in the end the Fermi theory had to be formulated in a field theoretical frame as a real weak interaction (which, I must admit, is what was going to happen in a couple of years). In such a framework, our consistency condition had no real justification.

However, the need of two different light quark masses has remained in the perturbative theory. Weak interactions produce divergent mass corrections which are different for the $up$ and $down$ quarks, leading to a priori different renormalised quark masses, to be determined form the physical violation of chiral symmetry, as we had done in our paper~\cite{CabibboMaiani1970a} and derived later by Steve Weinberg in the electroweak gauge theory~\cite{Weinberg1973}.

So, we discussed and discussed, apparently getting nowhere. Usually we were two of us, often in different compositions, arguing against the one who was at the blackboard. But during our discussions a change in paradigm occurred. 

Previous works had been done in the framework of the ``algebra of currents'', a rather formal and clumsy framework. But slowly we began to phrase our discussion in terms of quarks, a language in which you could associate concepts, more transparently, to Feynman diagrams, quark masses and the like. 

A second step was to realise that the previous arguments about rotating away parity and strangeness violation could not help with the Ioffe-Shabalin problem, with which the issue of the Cabibbo angle had nothing to do.

\begin{figure}[ht]
\begin{center}
\includegraphics[scale=.47]{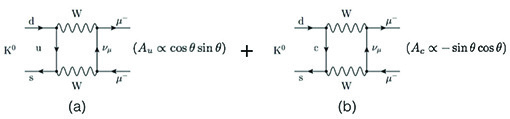}
\caption{\footnotesize{(a) Feynman diagram for the process: $K^0=(d{\bar s})\to \mu^+\mu^-$ according to Ioffe and Shabalin; the quark $u$ couples to $d_C=\cos\theta d+ \sin\theta s$ thus $A_u$ is proportional to $\cos\theta \sin\theta$. The GIM proposal is represented by (a)+(b); in (b) the quark $c$ couples to $s_C=-\sin\theta +\cos\theta$, $A_c$ is proportional to $-\sin\theta \cos\theta$ and the leading divergent amplitudes cancel (GIM mechanism).}}
\label{IS}
\end{center}
\end{figure}

In quark language, the Ioffe-Shabalin problem for $K^0\to \mu^+\mu^-$ was represented by the diagram in Fig.~\ref{IS} (a). The divergent amplitude is proportional to the product $\cos\theta \sin\theta$ of the couplings of the quarks $d$ and $s$ to the $u$ quark, as obtained from the Cabibbo theory (see {\bf Box 1}).

We turned around the Ioffe-Shabalin problem until January 1970, when we got convinced that the weak interaction theory had somehow to be modified. At this point, the solution was just under our eyes. 

A fourth quark of charge $+2/3$, coupled in a doublet to the $s_C$ quark left out by Cabibbo, had been introduced few years before by Bjorken and Glashow, for entirely different reasons, and had been called the {\it charm quark} \cite{Bjorken1964}. 

The exchange of the charm quark would provide an additional diagram whose divergent part had to be proportional to $-\sin\theta \cos\theta$, so that it could exactly cancel the divergent part of the Ioffe-Shabalin diagram, Fig.~\ref{IS} (a)+(b) (this has been called later the GIM mechanism). 

If the additional quark has a different mass from the up quark, it is simple to see that the sum of the two diagrams gives a {\it finite} result, which has exactly the Ioffe-Shabalin form with 
\begin{equation}
\Lambda^2\to m_c^2-m_u^2\nonumber
\end{equation}
{\it The anomalously low cutoff of $3$~GeV was the order of magnitude of the mass of the particles containing the charm quark}!

The latter result was unexpected and gratifying. The charm quark of Glashow and Bjorken was supposed to be much lower in mass and as such it had been already excluded by experiments. On the other hand, a quark with $~2$~GeV mass was perfectly compatible with not having been seen yet. As we would write later in our paper \cite[p. 2]{GlashowIliopoulosMaiani1970}:

{\it Why have none of these charmed particles been seen? Suppose they are all relatively heavy, say 2 GeV. Although some of the states must be stable under strong (charm-conserving) interactions, these will decay rapidly ( $10^{-13}$sec) by weak interactions into a very wide variety of uncharmed final states (there are about a hundred distinct decay channels). Since the charmed particles are copiously produced only in associated production, such events will necessarily be of very complex topology, involving the plentiful decay products of both charmed states. Charmed particles could easily have escaped notice.}\footnote{Starting from 1971, emulsions experiments performed in Japan by K. Niu and collaborators~\cite{Niu1971} did show cosmic ray events with {\it kinks}, indicating long lived particles (on the nuclear interaction time scales), see later, {\bf Box 4}.}

The resulting scheme is reproduced here
\begin{eqnarray}
&& \left(\begin{array}{c} u \\ d_C \end{array}\right)_L, u_R,d_R;~\left(\begin{array}{c} c \\ s_C \end{array}\right)_L, c_R,s_R\nonumber \\
&&\left(\begin{array}{c} \nu_e \\ e \end{array}\right)_L,e_R;~\left(\begin{array}{c} \nu_\mu \\ \mu \end{array}\right)_L,\mu_R.
\nonumber
\end{eqnarray}
It restores an exact quark-lepton symmetry for the left-handed doublets that participate in the weak interactions. 

This scheme had been considered, on purely aestethical grounds, by different authors, even before Glashow and Bjorken, see {\bf Box 4}. 

The cancellation of divergences was the signal of a most important property of the two quark doublet theory. Unlike what happens with Cabibbo's three quark theory, the commutator of the charged currents produces a \textit{neutral current without strangeness change}. Thus, it becomes possible to write a Yang-Mills theory of the weak interactions, or of unified weak and electromagnetic interactions, without incurring in the strangeness changing neutral current which had precluded the extension to quarks of the unified theories formulated by Glashow in 1961 \cite{Glashow1961} and by Weinberg and Salam in 1967-1968 \cite{Weinberg1967} \cite{Salam1968}.

We stated this point very clearly in the introduction of the paper:

 \textit{\dots\ we observe\dots\ that an extension to a three-component Yang-Mills model may be feasible. In contradistinction to the conventional (three-quark) model, the couplings of the neutral intermediary --- now hypercharge conserving --- cause no embarrassment. The possibility of a synthesis of weak and electromagnetic interactions is also discussed}. 
 
By the end of January, I think we had understood all the essentials and we were very happy.

I remember one day going to the Legal Sea Food for lunch where my wife Pucci  joined us. Pucci told to Shelly how happy and excited I was about the new result and the work we were doing. He replied: ``He is right, this paper is going to be on all school books.'' Shelly was fantastic\dots \ 

I remember another occasion, a seminar which Shelly gave to the experimentalists of Harvard, working at the CEA (Cambridge Electron Accelerator), and he said: ``Look, with charm we have essentially solved particle physics. Except'', he added, ``for the problem of $CP$ violation''. 

In fact we had tried to introduce $CP$ violation in our scheme, but we easily proved that, with two left-handed quark doublets, you could eliminate any complex phase from the weak hamiltonian: our theory was exactly $CP$ invariant. 

Unfortunately, we did not ask ourselves what would happen if we had another doublet, as Makoto Kobayashi and Toshihide Maskawa  would do three years later \cite{KobayashiMaskawa1973}. In 1970 I think we had the prejudice that one could not introduce CP violation with only left-handed interactions. The prejudice persisted, at least with me, until 1975, when I discussed the matter with Glashow in the school at Gif-sur-Yvette, and I seem to remember that the prejudice was shared by Shelly as well.

Before writing the paper, we wanted to talk with Francis Low, who had been one of the initiators of the story.
So we took an appointment and  all three went to MIT to see him. 

When we arrived in his office, Francis Low was speaking with Sergio Fubini and Gabriele Veneziano who, at the time, were at MIT formulating the dual model of strong interactions. However, they left shortly after Shelly started illustrating our ideas.

At this point, Steve Weinberg came in (I did not know Weinberg's paper at that time, I read it few days after this conversation). Weinberg started debating with Shelly and he was very negative. John and I did not intervene into these discussions between old time school friends, but assisted with amusement. Steve pointed out his arguments and Shelly responded but the conversation went on without getting to any conclusion. 

What I understood of Weinberg's view was that he believed the strong interactions would destroy the simple relations we had found with the free quark model --- the GIM mechanism. He said that in a complicated way and we could have pointed out easily ``But look, if strong interactions are with a neutral gluon \textit{\`a la} Preparata-Weisberger, they will not change the commutation relations.'' We had the answer and I am sorry we did not put it in the discussion! 

But the surprising point is that he did not say: ``Look, I have a theory\dots''. Because we said: ``With this theory you can do a unified theory also with quarks.'' He did not say: ``I have a theory with the electrons and could not extend it to the quarks because of strangeness changing neutral currents.'' He did not mention it. Then we left and decided to publish the paper anyway \cite{GlashowIliopoulosMaiani1970}. 

At that same time, Nicola was visiting Princeton, where he had to stay for a semester. So I decided that I would go and visit him and ask for his opinion. Pucci and I took a bus from Boston to Princeton and I told Nicola about the paper. I found that he, too, was not so enthusiastic. I was in a difficult position because our paper was going in a completely different direction as the one that Nicola and I had done on the angle. 

He was not very happy because, in a way, he didn't really understand the story. My interpretation is that he thought that we were cheating. I mean, to solve the problem we had introduced another quark, while the rules of the game, usually, are to solve the problems with the theory that you have and not by changing it. 

Anyway, discussing with Nicola was, as usual, very enlightening and we made an interesting reasoning. 
We had computed the weak angle by means of the consistency condition I mentioned before. How would the condition look like with four quarks? We quickly arrived to the conclusion that the four-quark scheme would {\it evade} the condition, as we use to say in the physicists jargon, meaning that the model did satisfy trivially the condition, for any value of the angle. This is true still today, in the Standard Theory, which includes the four-quark scheme. 

The empirical relation found by Gatto and by us between the Cabibbo angle and the ratios of quark masses, an impressive regularity, is still there, crying for a rational explanation.

The day after we had to leave, with Nicola informed but not convinced that one could not find another solution to the Ioffe-Shabalin problem. It was only in 1974, four years later, that he accepted that charm was inevitable. 

In early spring 1970, Pucci and I went to New York where I had to give  a seminar on our work at the invitation of Giuliano, who was at the Rockefeller Institute. Giuliano had a very violent reaction, attacking me at the end of the seminar, and considered our results unjustified. 

It had not to be the last attack, charm was received very coldly by our community, to say the least. But the attack of Giuliano left me troubled, I should say furious, given our friendship and the many works done together.

Then we arrive at Easter 1970 when all of a sudden the pregnancy of Pucci, then at the eighth month, stopped and she lost the baby we had cherished all the time in Cambridge and were preparing to receive.

The impact on Pucci and me was very violent, for the loss and for worries about the future. We were so upset that we decided to interrupt our stay in Cambridge and come back to Rome. It was not so easy to convince the landlord of the apartment and Harvard to resolve our contracts, but finally everybody understood.

The family of my wife was against  flying (for this reason, they never came to visit us during our stay in the US).  Her father, who was an influential journalist in Italy, found us a first class ride on the Cristoforo Colombo ocean liner. Shelly said: ``Well, if you have a first class cabin, you have to make a party!'' So we organised a party on board before the ship left:  we ordered sandwiches and Shelly arrived with a bucket filled with ice and champagne bottles. Other persons present were Sam Ting, John Iliopoulos and Paul Martin. Champagne was so abundant that our guests had some trouble in getting safely out of the ship, when departure was announced.

It was a very lovely party,  we were grateful for the love shown to us by our friends. John took some pictures, I have seen one but, apparently, he has not been able to retrieve any of them, at the time of this writing. 

Ten days later, after a very gratifying travel, we arrived in Naples, with the whole family waiting for us in the harbour. Thus we settled again in Italy. 

\section{Back to Italy}

 L. B. \hspace{0.2 cm}
In the meantime the $e^+ e^-$ collider Adone had gone into operation at Frascati Laboratories\dots 
 \vskip 0.3 cm

L. M.  \hspace{0.2 cm}
And in fact, at my arrival in June 1970, I found Nicola working with Giorgio Parisi and Massimo Testa on the preliminary results of Adone. They were very excited and wrote an important paper on the parton model for $e^+ e^-$ annihilation \cite{CabibboParisiTesta1970}.

Bruno Touschek was also very excited by the abundant hadron production found by Adone \cite{BacciEtAl1972} \cite{BacciEtAl1973}. 
 
Now, the interesting point of our paper was not only the GIM mechanism, but also the fact that the Ioffe and Shabalin calculation of the cutoff gave an order of magnitude of the mass of the particles containing the charm quark. So we had a reasonable guess of where one should find them and this energy was very close to the maximum energy that Adone could reach, which was in fact precisely $3$~GeV.  

I talked with various people about searching charmed particles as well as the $c\bar c$ vector particle, the analog of the $\phi$ meson (later discovered, in 1974,  by Ting and Richter as $J/\Psi$ at $3.1$~GeV).

I remember, in particular, talking with Ugo Amaldi at Istituto di Sanit\`a and discussing with him the experimental signatures of charm. Ugo took the proposal seriously, in particular the suggestion that neutrino interactions could produce charmed particles which, after successive semileptonic decay, would give rise to events with a pair of opposite charge muons. Ugo would come back to this circumstance later, when he led CHARM (CERN, Hamburg, Amsterdam, Roma, Moscow Collaboration), one of the second generation CERN neutrino experiments. 

However, it was really out of the mind of people that there could be another spectroscopy beyond strange particles and nobody payed any attention. Even more, they started looking at me like one of those foolish people that tell you that there will be new phenomena, while everything is normal, there are only 3 quarks, and everybody knows that. 

So, in fact, I did not work on the GIM paper any longer, waiting for times to become ripe for new discoveries. In Harvard, John and Shelly worked on the massive Yang-Mills theory (i.e. without spontaneous symmetry breaking) to see whether it could be more convergent or even renormalisable. They proved many cancellations of potential infinities but did not come to any conclusion \cite{GlashowIliopoulos1971}.

Shelly rates this as the most useless paper that he ever wrote, but  it contains a lot of ingenuity and they must have had a lot of fun. 

Shelly and John went for summer vacations in Mexico, visiting CINVESTAV, the Center for Research and Advanced Studies of the National Polytechnic Institute. In Mexico there was the football World Cup. They went to see the semifinal, Italy against Germany, at the Azteca stadium, and wrote me a postcard celebrating the victory of ``the Italians'' (unfortunately, Italy lost the final to Brazil). 

To close this period, I recall the Italian Physical Society conference of November 1970, in Venice. I was asked to give a talk and presented the GIM work.

One of our suggestions was  that there had to be semihadronic, strangeness conserving, neutral currents that had to be seen as neutrino interactions on nuclei without a final muon or electron (these events were to be discovered in 1973 by Gargamellle at CERN).

I made this talk and, to my surprise, Nino Zichichi had an energetic reaction, saying that theorists should not lose their time in these idle speculations: ``There are problems that theorists should better do, like studying the $\omega-\phi$ mixing'' and the like.  He was quite aggressive. 

At that point, Nicola intervened. He did not say: ``This is an interesting paper,'' because he, too, was not convinced, but he simply observed: ``They made a proposal, they made predictions, and we will see whether these predictions are right or wrong. Why should you complain?''.

Nino's reaction was perhaps a consequence of the talk I had given in summer in Erice. There was Giuliano Preparata and my talk had not been very well received. 

These episodes show how badly the idea of charm was received by everybody, from Nicola to Preparata to Nino Zichichi\dots \ It was really considered as a nuisance or, in the least, a trick of dubious taste.

\section{New Discoveries}


L. B. \hspace{0.2 cm} At the beginning of the 1970s, Particle Physics seemed to be heading towards dual (or string) models of the hadrons. What happened then?

\vskip 0.3 cm

L. M.  \hspace{0.2 cm}
Yes, dual models dominated the scene. However, the years 1971-1973 brought decisive discoveries.

In 1971, 't Hooft and Veltman showed the renormalisability of the Weinberg-Salam theory \cite{tHooftVeltman1972}. 

In 1972, Bouchiat, Iliopoulos and Meyer proved the cancellation of Adler anomalies in an Electroweak theory  with quark-lepton symmetry and fractionally charged quarks in three colors \cite{BouchiatIliopoulosMeyer}.

In 1973 there was the discovery of neutral currents by Gargamelle at CERN \cite{HasertEtAl1973a} \cite{HasertEtAl1973b}\dots

\dots and in the same year came the discovery of asymptotic freedom of the Yang-Mills theory by Gross and Wilczeck \cite{GrossWilczek1973a} \cite{GrossWilczek1973b} \cite{GrossWilczek1974} and by Politzer \cite{Politzer1974a} \cite{Politzer1974b}. 

Shortly after, the idea of color interaction of quarks was put forward by Fritzsch, Gell-Mann and Leutwyler \cite{FritzschEtAl1973}.

 In three years, the paradigm shifted completely towards field theory, a shining example of what Thomas Kuhn in 1962 had called a {\it scientific revolution} \cite{Kuhn2012}. What came to be called the \textit{Standard Theory} took form.

But let me go in order, starting from the Amsterdam International Conference on Elementary Particles, Amsterdam 1971.

In 1971, Shelly was in Europe, visiting Marseille. I went to visit him (we had a wonderful fish dinner in Bandol) and  then we went together to the conference  in Amsterdam, where I spoke about some aspects of the spontaneously broken, $SU(3)\otimes SU(3)$ chiral symmetry \cite{AltarelliEtAl1971}.
I was given as scientific secretary a very young and very kind PhD whose name was Gerard 't Hooft! This is how I met Gerard.  

That summer, before the conference, Veltman had  made a seminar in Marseille, announcing the renormalisability of the Weinberg-Salam theory. They considered Weinberg-Salam with leptons only.

Adler's anomalies in $SU(2)\otimes U(1)$ were the last obstacle towards a renormalisable electroweak theory. In a very important paper, Claude Bouchiat, John Iliopoulos and Phil Meyer proved that the anomalies could be cancelled between quarks and leptons \cite{BouchiatIliopoulosMeyer}. John's description of this work, in a short letter sent to me immediately after: ``there must be charm, quarks have color and are fractionally charged."

At that point charm went on the road. In 1972, at the ICHEP Conference held at FermiLab, Ben Lee presented the electroweak theory with leptons and quarks extended to charm. It was the bifurcation point where the GIM mechanism and charm were accepted by theorists as part of the standard lore.

In 1973,  at  the 2nd  Aix en Provence International Conference on Elementary Particles, the discovery of muonless neutrino events signaled the existence of neutral weak currents and the beginning of a new era (for a more detailed historical reconstruction see~\cite{Haidt2004}). 

Before the Gargamelle papers were published, in spring 1973, Ettore Fiorini had called me in Rome saying he wanted to discuss with me and Nicola about new unexpected results with Gargamelle. He came and we spent a whole afternoon listening to him about neutrino interactions without a final muon, asking questions about possible backgrounds and trying to find weak points in his story. It looked all very solid to us. 

I remember waking up the day after and thinking back to what Fiorini had told us. While shaving, I looked myself in the mirror and said to myself: ``Look, we have understood everything. Now we know everything.'' I was like Shelly was three years before (and, similarly, with CP violation still unsolved).  


With three colors, one could have  color singlet baryons made of three quarks obeying Fermi statistics. Asymptotic freedom guaranteed approximate scaling in deep inelastic and deep $e^+ e^-$ annihilation. Electrically neutral gluons made so that strong interactions did not interfere with electroweak renormalisation, including GIM mechanism. A complete theory of the strong, electromagnetic and weak interactions with two quark and lepton doublets was there!

Of course, charmed particles and intermediate vector bosons {\it had} to exist.
Later, commenting the neutral currents discovery, I remember telling Giorgio Salvini: ``Look, the best has yet to come, yet to come!''. There would have to be many other surprises\dots

The ICHEP conference in 1974 was held in London. John Iliopoulos gave the talk on unified theories \cite{Iliopoulos1974}, a wonderful talk, saying that we had understood everything of the old problems and were now aiming at the grand unification of all interactions. It was a great triumph, concluded with a public bet by John of a bottle of good French wine that charm would be discovered in two years. 

There was still the issue of CP violation, but nobody really mentioned it.  

\section{Octet Enhancement}

L. B. \hspace{0.2 cm} Coming back to the Weak Interactions, could one say  at that time that everything had been understood? 

\vskip 0.3 cm
\noindent L. M.  \hspace{0.2 cm} Studying QCD when it was still the new sensation, Guido Altarelli and I came back to an old problem of weak interaction physics, which had raised a lot of attention in the 1960s, particularly in Florence. This was the problem of the so called $\Delta I=1/2$ rule (see {\bf Box 3}). In the Eightfold Way of Gell-Mann and Ne'e man, it had become the problem of {\it octet enhancement}, but it had not been anyway possible to explain the mysterious ``deformation'' of the product of two Cabibbo currents that amplified the $\Delta I=1/2$ over the $\Delta I=3/2$ component.

With asymptotic freedom, it was possible, for the first time, to use perturbation theory to compute reliably the anomalous dimension of operators such as the non-leptonic hamiltonian. Could we give substance to the old idea of Ken Wilson \cite{Wilson1969} that the enhancement was due to a renormalization effect of the product of the currents due to strong interactions?

It took some time to me, in winter 1973, to get into the formal part of the calculation. As I said before, I did not spend much time in Florence to study renormalisation theory and I had to come back to my basic. 

One physics question puzzled us at the beginning. How can flavor-blind QCD tell isospin 1/2 from isospin 3/2?
 The answer came from an old observation of Richard Feynman, who had noticed that if quarks were bosons the Fermi non leptonic interaction would be pure $\Delta I=1/2$.
 The reason has to do with the exchange of $u$ and $d$ quarks in the non-leptonic hamiltonian. Quark fields carry Dirac indices multiplied to a definite product of Dirac matrices.  To see the effect of exchanging $u$ and $d$, we have to take into account the statistics of the fields (bosons commute, fermions anticommute), and also the effect of the exchange of the Dirac indices. With the $V-A$ interaction, exchange of the Dirac indices gives a factor $-1$. If quarks were bosons, the exchange of fields would give $+1$ and total exchange $-1$. With $u$ and $d$ antisymmetric, the $ud$ pair is in pure isospin $I=0$ and the hamiltonian has pure $\Delta I=1/2$. This was Feynman's observation.
  
 With color, quark fields are given a new index, saturated with appropriate, color invariant, matrices and we have to take into account the effect of exchanging color indices as well. It is seen that we may have both symmetry and antisymmetry (see Box 3). The antisymmetric case would correspond to: $-1$ for Dirac, $-1$ for color, $-1$ for Fermi statistics, in total $-1$ for the exchange and $\Delta I =1/2$. The antisymmetric and symmetric combinations behave differently under color renormalisation and the enhancement of the pure $\Delta I=1/2$ component is possible.

\vskip 0.3 cm 
\begin{tcolorbox}[breakable, enhanced]
\footnotesize{
{\bf Box 3. $\Delta I=1/2$ Enhancement in Non-Leptonic Decays}\\
\vskip 0.1cm
$\Delta I=1/2$ enhancement is a prominent feature of non leptonic decays of strange particles.
The product of the Cabibbo currents for $d\to u$ (I=1) and $s\to u $ (I=1/2) should lead to a balanced mixture of 1/2 and 3/2, while the rate of $K_S  \to \pi^+ \pi^-~(\Delta I=1/2)$ is larger than the rate of $K^+\to \pi^+ \pi^0,~ (\Delta I=3/2)$ by about a factor of $400$.

In 1969, Ken Wilson had noted \cite{Wilson1969}  that the strong interactions, which respect Isospin conservation, could renormalise differently the two components. However, without a theory of the strong interactions he could not test the idea. But what about  QCD?

In formulae, the non-leptonic hamiltonian is (see Box 1 for the Cabibbo currents, no color):
\begin{eqnarray}
&& H_{non-lept}= \frac{G_F}{\sqrt{2}}[{\bar s}_L \gamma^\mu u]~[{\bar u}_L \gamma_\mu d]~+~{\rm h.c.}= \frac{G_F}{\sqrt{2}}[({\bar s}_L)_\alpha \gamma^\mu_{\alpha \beta} u_\beta]~[({\bar u}_L)_{\delta}( \gamma_\mu)_{\delta \epsilon} d_{ \epsilon}]~+~{\rm h.c.}\nonumber
\end{eqnarray}

The isospin of the operator is obtained from its symmetry properties under the exchange $u\leftrightarrow d$, with $\bar s,~\bar u$ fixed. Antisymmetry under exchange corresponds to the pair $ud$ in isospin zero, therefore pure $\Delta I=1/2$, while the symmetric combination has $\Delta I=3/2$ and $1/2$ in similar proportions. 
In $SU(3)$, $ud$ antisymmetry corresponds to pure octet and $ud$ symmetry to a mixture of octet and decuplet.

With the $V-A$ interaction, exchange of the Dirac indices $\beta$ and $\epsilon$ gives a factor $-1$. If quarks were bosons, the exchange of fields would give $+1$ and total exchange $-1$. This is Feynman's observation.

With color, quarks acquire new indices $i,j,\cdots$, saturated with appropriate, color invariant, matrices (color singlets weak currents, Dirac indices ignored):
\begin{eqnarray}
&& H_{non-lept}= \frac{G_F}{\sqrt{2}}[{\bar s}_L \gamma^\mu u]~[{\bar u}_L \gamma_\mu d]~+~{\rm h.c.}= \frac{G_F}{\sqrt{2}}[{\bar s}_L^i \gamma^\mu u_k]~[{\bar u}_L^j \gamma_\mu d_{ h}] \times \delta^k_i \delta_j^h~+~{\rm h.c.}\nonumber
\end{eqnarray}
and we have to take into account also the effect of exchanging color indices $k,h$ in the color matrices. We may have both symmetry and antisymmetry
\begin{eqnarray}
\delta^k_i \delta_j^h= \frac{1}{2}(\delta^k_i \delta_j^h+\delta^h_i \delta_j^k)+ \frac{1}{2}(\delta^k_i \delta_j^h-\delta^h_i \delta_j^k)\nonumber
\end{eqnarray}

The antisymmetric case would correspond to: $-1$ for Dirac, $-1$ for color, $-1$ for Fermi statistics, in total $-1$ and $\Delta I =1/2$. The antisymmetric and symmetric combinations behaving differently under QCD renormalisation, enhancing the pure $\Delta I=1/2$ component is possible.

By asymptotic freedom, we may assume  the non-leptonic hamiltonian, at momentum $M_W$, to be approximately equal to the product of the Cabibbo currents for $d\to u$ and $u \to s$. Gluon exchange, at lower momenta, would renormalise differently the $\Delta I=1/2$ and $3/2 + 1/2$ components, in a way computable in perturbation theory. Integrating the renormalisation group equations from $M_W$ to the typical momentum exchanged in non-leptonic decays of strange particles, $\mu$, one expects enhancement or suppression factors of order $(M_W/\mu)^d$, where $d$ is the so-called anomalous dimension. With the scale of K decays $\mu << M_W$, the enhancement could be sizeable for the component of the Hamiltonian which has $d>0$ (in our paper, $M_W \sim 60-100$ GeV was assumed, taken from the Weinberg-Salam theory and the value of the Weinberg angle from neutrino neutral current data). 

The calculations in \cite{GaillardLee1974} \cite{AltarelliMaiani1974}  have shown that, indeed, the $ud$ antisymmetric component with $\Delta I=1/2$ has $d>0$ and the symmetric one, with $\Delta I=3/2 + 1/2$ has $d<0$.
}
\end{tcolorbox}
\vskip 0.3 cm


It did not take much effort to us to compute gluon exchange in the four quark non-leptonic hamiltonian (a calculation very similar to the radiative corrections to $\beta$ decays done with Nicola and Giuliano in the 1960s). To our delight, we found a positive  anomalous dimension for the $ud$ antisymmetric hamiltonian and negative dimension for the symmetric component.

\begin{figure}[ht]
\begin{center}
\includegraphics[scale=.45]{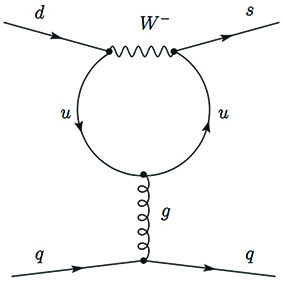}
\caption{{\footnotesize Penguin diagram.}}
\label{peng}
\end{center}
\end{figure}

At this point, however, we found another operator that can mix, see Fig.~\ref{peng}. We had discovered what was later called a {\it penguin diagram} by John Ellis.

It took some time for us to understand if this contribution was correct or not, we had to use the equation of motion of the gluon to reduce the penguin to a four-quark operator and if that was legal was not clear to us. The positive thing was that it was a pure $\Delta I=1/2$ operator and it had a positive dimension, even larger that the dimension of the antisymmetric four-quark hamiltonian.

Finally, we accepted the penguin and we rushed out a preprint with the larger anomalous dimension. 

Soon after, we received a  paper by Ben Lee and Mary K. Gaillard, with a similar title and the same physics but without mention of the penguin.

We were puzzled, but soon I went to the London Conference where I saw Mary K. I went directly to ask her what they did about the penguin. She looked at me with surprise and said: ``You know...GIM mechanism...''. Adding charm quark exchange in the penguin simply eliminated its contribution to the anomalous dimension.

I called Guido to inform him and we quickly sent to \textit{Physics Letters} a revised version of our paper \cite{AltarelliMaiani1974}, which came out at about the same time as Ben and Mary K. paper \cite{GaillardLee1974}.

I was very ashamed, and still today cannot understand why I did not consider the charm quark. The only explanation is the surprise at the appearance of the penguin and the effort to give it a meaning.

However, still at the London Conference, I met Ken Wilson and was happy to tell him that QCD supported his old conjecture on the origin of the $\Delta I=1/2$ rule. This was in fact the first success of QCD outside deep inelastic scattering and it was widely publicized.

Had we considered a strangeness not changing transition, charm and up quark exchange  would add and one would have to consider penguins and four quark operators all together. This was the case of Parity violation in atomic physics that Marie-Anne and Claude Bouchiat had  started investigating at \'Ecole Normale in Paris (in that case one had to consider {\it electroweak penguins}, with the gluon replaced by a photon or a $Z^0$). We made a point of honor to start an analysis of this case, involving our young postdocs  Roberto Petronzio and Keith Ellis \cite{AltarelliEtAl1975}.

More important, few years later penguins reappeared in the $\Delta I=1/2$ story, with a beautiful paper from ITHEP (Moscow) \cite{ShifmanEtAl1977}.

One could divide in two regions the integration from $M_W$ to the K-decay momentum scale: above and below  the charm mass, $m_c$. In the lower region, charm cannot be excited and one finds the penguin of Fig.~\ref{peng} alone. This region has a lower span in momentum, one gets effects of order $(m_c/\mu)^d\sim (m_c/m_s)^d\sim 10^d$  ($m_s$ is the strange quark mass) rather than $(M_W/ms)^d$, but the strong coupling constant, which appears in the anomalous dimension $d$, is larger and the penguin is important. In this way, combining the results of the two regions, SVZ could get an estimate of the enhancement somewhat closer to the physical value. 

\begin{figure}[ht]
\begin{center}
\includegraphics[scale=.75]{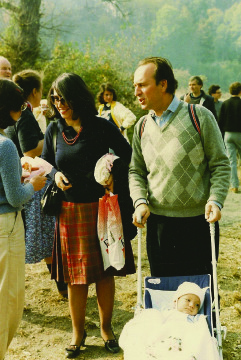}
\caption{{\footnotesize With Pucci and Camilla, at the Theory Division picnic, CERN 1985.}}
\label{thpicnic1985}
\end{center}
\end{figure}

However, theoretical estimates in the lower region are more uncertain, since, opposite to asymptotic freedom, strong interactions are becoming stronger and perturbative estimates less reliable.

The question if QCD can explain \textit{quantitatively} the $\Delta I=1/2$ rule will be finally answered by high precision, non-perturbative lattice QCD calculations in the low momentum region. 

In the years 1980s, in collaboration with Belen Gavela and Olivier P\`ene, Orsay, and Chris Sachrajda, Southampton,  we have given several contributions, both theoretical \cite{BochicchioEtAl1985} \cite{MaianiEtAl1987} \cite{CurciEtAl1988} and computational \cite{GavelaEtAl1988} \cite{GavelaEtAl1989}, without reaching, however, the needed numerical precision.

As far as I can tell, this is still an open issue.

\section{Charm, at Last!}

After the London Conference, there was the Gif-sur-Yvette School \cite{CabibboEtAl1974}.

Nicola  and Guido were both there. Nicola had gotten finally convinced about charm and we concluded that we had to tell people in Frascati to search seriously for a new vector meson. A meeting in Frascati had already been scheduled, but the day before Ting announced the discovery of the $J$ particle in Brookhaven and so did Richter for the $\Psi$ particle seen at SLAC with the same mass. It was the  November revolution! 

Frascati understood the opportunity they had lost and immediately set to work to increase the energy some $100$~MeV above the maximum design energy, to be able to see the $J/\Psi$ particle, as it was called not to displease anybody.  

Everybody presented his/her own theory about the new particle.  We thought immediately to the $c\bar c$ vector mesons, predicted by GIM around a mass very consistent with the $J/\Psi$ mass of $3100$~MeV. 

However, there was something strange about the charm interpretation, in regard of the $J/\Psi$ width. GIM had estimated the width to be small because, most likely, the expected vector meson was below the threshold to decay into a pair of charm-anticharm mesons. Our guess for its decay into normal hadrons was derived from $\phi$ decay into three pions, and was of the order of MeV. The observed width was $100$ keV, a factor $10$ times smaller: why was that?

In addition, with the $J/\Psi$ produced and observed in Frascati, rumors started to leak from the experimental collaborations. In particular, we got from one of the experiments that there was some evidence for a backward-forward asymmetry in the decay $J/\Psi \to \mu^+ \mu^-$.

The latter was a well known signal of parity violation, studied long before by Gatto and Cabibbo as the typical signature of a neutral intermediate vector boson (IVB). So we started wondering if the new particle was really $c\bar c$ and, if not, what was it? Was it perhaps a light neutral IVB?
  
 The most direct check was with the $\mu^+\mu^-$  width, which, for the neutral IVB is directly related to the Fermi constant and to the particle mass. The mass and the muonic width were already known and the calculation of the ``would be Fermi constant'' could be made on the back of an envelope. So, we came to one of those incredible coincidences that, alas, happen from time to time. The result was within $20\%$ equal to the true Fermi constant.
 
So, on the Saturday morning following the first Adone run we wrote a paper \cite{AltarelliEtAl1974a}. We had to call Elisabetta Disilvestro, secretary to the Director, to come to the Institute and make the manuscript into a decent preprint. Then we sent Roberto, the junior member of the group, to Bologna by train, to give the typewritten paper to the Editorial Office of \textit{Nuovo Cimento Letters}.

Then I was invited by Salam to go to Trieste to make a seminar at ICTP on the paper with Guido on Octet Enhancement. In Trieste, I got a phone call from Nicola who said: ``We are wrong. They have seen at SLAC another particle, the $\psi '$, so  it is not the $Z^0$ but a hadron''. In addition, he mentioned a paper by De Rujula and Glashow claiming that the asymptotic freedom of QCD would reduce the $J/\Psi$ width to 100 keV without problems \cite{DeRujulaGlashow1975}. The charm-anticharm hypothesis had been advanced, in the same days, by Cesareo Dominguez and Mario Greco \cite{Dominguez:1974be}, without however addressing the narrow width problem.

In fact, Harvard people were about one light-year ahead of us, due to a very clever paper made by Appelquist and Politzer \cite{AppelquistPolitzer1975} during summer. The idea was that a pair of very heavy quarks would be bound by color forces at distances of order of $1/m_c$, much smaller than typical hadron distances. Due to asymptotic freedom, these forces would be dominated by one-gluon exchange and the pair would form bound states very similar to positronium  (electron-positron pair bound  by one photon exchange forces), hence the name ``charmonium'' given to these states by the Harvard group \cite{AppelquistEtAl1974}. 

The  lightest charmonia would be below threshold for dissociating into a pair of charmed mesons and would decay into two or more gluons (similar to positronium decays into photons) which then would materialise into normal, uncharmed, hadrons. In particular the spin one state, the $J/\Psi$, had to decay into three gluons, which gave a substantially small width, compatible with the estimates of the strong coupling constant at momentum scales of the order of $1/m_c$ \cite{DeRujulaGlashow1975}.

An important information, besides the $J/\Psi$ width, was the so-called ratio $R$, the ratio of the multihadron to $\mu^+ \mu^-$ cross sections, a function of the center of mass energy:
\begin{eqnarray}
&& R(s)=\frac{\sigma (e^+ e^- \to {\rm hadrons})}{\sigma (e^+ e^- \to \mu^+ \mu^-)}, ~\sqrt{s}=2E~ \nonumber
\end{eqnarray}
with $E$ the center of mass beam energy.

At Frascati, $R$ was at a value around $2$ (with large errors, see Fig.~4 in~\cite{Richter1974}, see also the analysis in~\cite{Dominguez:1974be}). At SLAC, after the $J/\Psi$ and $\psi^\prime$ region, $R$ raised approximately by $2$ units.

After the work of Cabibbo, Parisi and Testa mentioned before, $R$ was analysed in terms of the production of pointlike partons. Assuming partons to be spin 1/2 particles, i.e. quarks, the formula was
\begin{eqnarray}
&& R(s)=\sum_i  Q_i^2\nonumber
\end{eqnarray}
with $i$ running over quark types. In three color QCD, $u,~d$ and $s$ quarks would give $R=3\cdot(4/9+2\cdot 1/9)=2$, which went well with the Frascati data. The threshold for $c\bar c$ production should then correspond to an increase $\Delta R=3\cdot 4/9\sim 1.3$.

In fact, there remained a lot of doubts wether the $J/\Psi$ could  be charm-anticharm,  first because the charm particles could not be found above the $J/\Psi$ and $\psi^\prime$ region. Secondly, because the increase of $R$ was about twice what expected.
So there were all sort of theories with other quarks or other. 

All that lasted until mid 1975 to 1976, when the $\tau$ heavy lepton was identified by Martin Perl and collaborators at SLAC. And it put all in order. 

First, $\tau$ production provided an $R$ increase of one unit, leaving the other unit to charm. Secondly, $\tau$ was the main source of the $e-\mu$ events that had kinematic properties that could not be identified with pairs of charmed particles. Also, in the same year, Lederman and collaborators discovered another series of narrow resonances, promptly identified with the lower member of another quark doublet, therefore called ``bottom'', a name promptly changed into ``beauty'' that sounded more appropriate: $\tau$ and $b$ where the signal of a new generation, a new complex with one quark and one lepton doublet.
 \vskip 0.3 cm

L. B. \hspace{0.2 cm}
 And this brings us to 1976, when at last, charmed D mesons were found\dots 
 \vskip 0.3 cm

\begin{figure}[ht]
\begin{center}
\includegraphics[scale=.20]{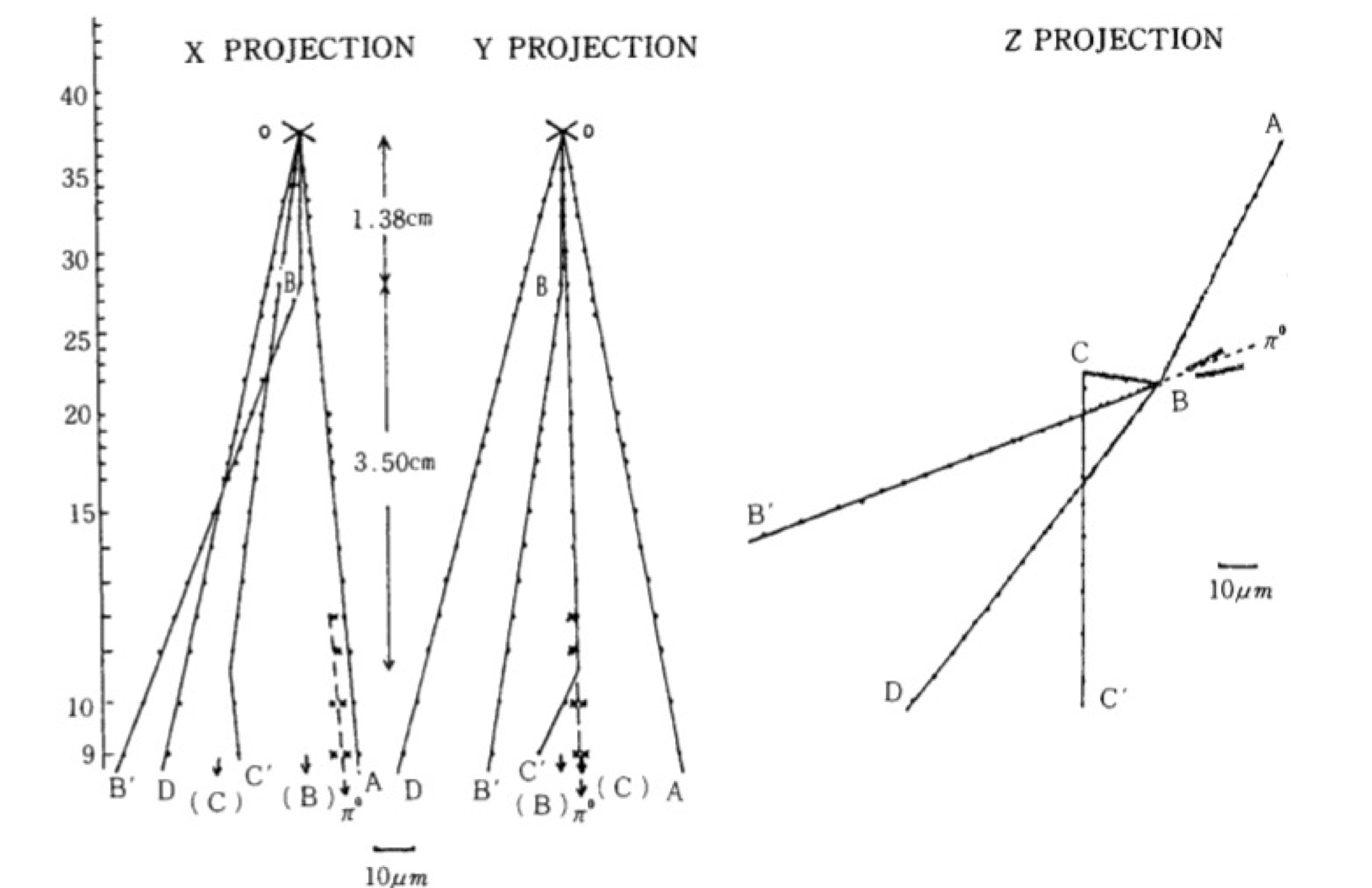}
\caption{{\footnotesize  {\it Pair production and decay of naked charm particles discovered in 1971 in a cosmic-ray interaction. Particle $B$ decayed at $B$ into $B^0$ and a  $\pi^0$. Two $\gamma$ rays, daughters of the  $\pi^0$, initiated electron showers at plate no. 12 and no. 10, respectively. Particle $C$ decayed at $C$ into $C^0$ and unseen neutral hadron(s)}. Figure and figure caption from~\cite{Niu2008}.}}
\label{Niu}
\end{center}
\end{figure}

L. M.  \hspace{0.2 cm}
Yes, by the Mark II Collaboration at SLAC, after a long search. And they decayed exactly as advertised in the GIM paper and further elaborated in a lucid paper by De  R\'ujula, Georgi and Glashow \cite{RujulaGeorgiGlashow1976}.
Nobody had doubts but Giuliano, then at CERN, who, to explain the narrow D, proposed that it could be a resonance with very high angular momentum. With charm, he had been always on the wrong side. 
\vskip 0.3 cm 
\begin{tcolorbox}[breakable, enhanced]
\footnotesize{
{\bf Box 4. A brief history of Charm}\\
\vskip 0.1cm
In the mid-1950s, the Sakata model \cite{Sakata1956} featured three basic constituents of the hadrons, (p, n, $\Lambda$), in parallel to the three elementary leptons known at the time:
\begin{eqnarray}
{\rm elementary \;hadrons} =\;\left(\begin{array}{ccc} & p &   \\ 
n &  & \Lambda\end{array}\right);\;
\;\; {\rm leptons}=\;\left(\begin{array}{ccc} & \nu &  \\e &  & \mu\end{array}\right)
\end{eqnarray}

In 1962, after the discovery of the muon neutrino, Sakata and collaborators \cite{SakataEtAl1962}, at Nagoya,  and Katayama and collaborators  \cite{Katayama1962}, in Tokyo, proposed to extend the model to a fourth baryon,  called V$^+$:
\begin{eqnarray}
{\rm elementary \;hadrons} =\left(\begin{array}{cc} p &V^+  \\  n & \Lambda\end{array}\right);\;
\;\; {\rm leptons}=\left(\begin{array}{cc}\nu_1 & \nu_2   \\ e & \mu \end{array}\right)
\end{eqnarray}
A possible mixing among  $\nu_e$ and $\nu_\mu$ was paralleled by $n-\Lambda$ mixing {\it \` a la} Cabibbo, giving rise to weak couplings of $p$ and $V^+$ entirely analogous to the couplings  $u-d_C$ and $c-s_C$ we  have considered in Sect.~\ref{gimdisc}.

In the seminal paper on quarks \cite{Gell-Mann1964}, Gell-Mann mentioned the possibility of a fourth constituent that would allow integral quark charges and restore the quark-lepton symmetry in the weak couplings, which were entirely analogous to those introduced in~\cite{SakataEtAl1962,Katayama1962}, an idea followed by Tarjanne and Teplitz \cite{Tarjanne1963} and by Hara \cite{Hara1963}.


In 1964, J. Bjorken and S. Glashow \cite{Bjorken1964}, reconsidered Gell-Mann's model with a fourth, integrally charged, quark and quark-lepton symmetric weak couplings.  They noted that, with the new addition, the commutator of two charged currents gave a flavor conserving neutral current, commenting that this {\it suggests a possible intimate connection between weak and electromagnetic interactions}. Mesons in~\cite{Bjorken1964} were assumed to be made by $q\bar q$ pairs and restricted to the irreducible $15$ dimensional multiplet of $SU(4)_{flavour}$. With this assumption, the mass of the $SU(3)_{flavour}$ singlet, made by a $c\bar c$ pair and identified with the newly discovered $\eta^\prime(958)$, was tied to the mass of the light  mesons and the masses of charmed mesons resulted to be of the same order than the $K$ mass. The existence of charmed mesons in this energy range was quickly excluded.


In 1970, there was  in fact no experimental evidence of weakly decaying hadrons beyond the lowest lying strange baryons and mesons. The fact that hadrons could be made with only three types of quarks was accepted as almost self evident. The theory of charm had to explain first of all why,  accelerator energies being already well above the mass scale indicated by the cutoff of Ioffe and Shabalin (see {\bf Box 2}), none of the charmed particles had been seen. This question was answered in \cite{GlashowIliopoulosMaiani1970} in terms of the large mass of charmed particles, at variance with the early ideas advanced in~\cite{Bjorken1964}, their pair production in hadronic collisions and their complex weak nonleptonic decays (see text). 


Starting from 1971, emulsions experiments performed in Japan by K. Niu and collaborators~\cite{Niu1971} did show cosmic ray events with {\it kinks}, indicating long lived particles (on the nuclear interaction time scales) with lifetimes in the order of $10^{-12}$ to $10^{-13}$  sec (see Fig.~\ref{Niu}). The lifetimes are in the right ballpark for charmed particles and indeed they were identified as such in Japan (see~\cite{Niu2008} for more details). However, cosmic rays events were not paid  much attention in western countries.

After 1972, with the proof of renormalisability of the Weinberg-Salam theory, charm and the GIM mechanism have been accepted as integral elements of electroweak unification. Higher order corrections leading to Flavour Changing Neutral Current (FCNC) processes have been computed in the renormalisable theory by M. K. Gaillard and B. W. Lee~\cite{Gaillard1974b}, and the consequences of a fourth quark extensively reviewed~\cite{Gaillard1974c}, based on the estimate of the charm quark mass, $m_c\sim 1.5$~GeV, derived in~\cite{Gaillard1974b}.

The first unequivocal evidence for a $c\bar c$ state was provided  in 1974 by the $J/\Psi$ particle ($M_{J/\Psi}=3.097$ GeV) discovered by C. C. Ting and collaborators \cite{Aubert1974} at Brookhaven, by B. Richter and collaborators \cite{Augustin1974} at SLAC and immediately after observed in Frascati \cite{Bacci1974}. The discovery came with the surprise that the $J/\Psi$ was much narrower than anticipated in  \cite{GlashowIliopoulosMaiani1970}. This was interpreted \cite{DeRujulaGlashow1975} as a manifestation of the {\it asymptotic freedom} at small distances of the color forces, which bind the quark-antiquark pair, recently discovered by D. Gross and F. Wilczek~\cite{GrossWilczek1973a} and by D. Politzer~\cite{Politzer1974a}. 
The charm quark is heavy enough for the $c\bar c$ pair to be separated by small distances such that color forces are already in the small coupling regime.  

Starting from 1973, {\it opposite charge dimuons} have been observed in neutrino high energy reactions. These events were interpreted to signal the production of a charmed particle that decays with the emission of a second muon. The process was:
\begin{equation}
\nu_\mu + N \to \mu^- + (c-{\rm containing~hadron})+\cdots\notag
\end{equation}
followed by the semileptonic decay:
\begin{equation}
c\to s +\mu^+ +\nu_\mu\notag
\end{equation}
which shows that the second muon has opposite charge to the first muon. In antineutrino reactions, the roles of the two muons are exchanged and the resulting muon charges are similarly opposite (see~\cite{Maiani1976b} for an early suggestions of this interpretation of opposite charge dimuons).

Open charm particles have been searched in $e^+ e^-$ colliders by the most visible signature, the so-called $e-\mu$ events, originated by the semileptonic decays of the lightest charmed particles:
\begin{eqnarray}
&& e^+ e^- \to c+\bar c \to e^+ ({\rm or} \; \mu^+) + \mu^- ({\rm or} \;e^-) + {\rm hadrons}
\end{eqnarray}

Events of this kind were observed by M. Perl and collaborators \cite{Perl1975}  at energies above the $J/\Psi$, but with the wrong energy distribution of the leptons to be produced in charm beta decays. A state of confusion ensued, until it was realised that the $e-\mu$ pairs were being produced by pairs of entirely unexpected new particles, which also decayed semileptonically. These were pairs of the heavy lepton $\tau$, whose threshold, for yet unexplained reasons, happens to be quite close to the $c\bar c$ threshold. 

It was only in 1976 that this fact was clearly recognised. The multihadron events in $e^+ e^-$ annihilation, after subtraction of $\tau$-pairs events, showed clearly the $c\bar c$ threshold with the jump in the cross section of the size corresponding to a spin 1/2, charge 2/3, particle.

The lightest weakly decaying charmed meson, $D^0 = (c\bar u)$ ($M_{D^0}=1.865$ GeV) was discovered in 1976 by the Mark I detector \cite{Goldhaber1976} at the Stanford Linear Accelerator Center. The charged meson $D^+=(c\bar d)$ ($M_{D^+}=1.870$ GeV) and the lowest lying baryons, $\Lambda^+_c =[c(ud)_{I=0}]$ ($M_{\Lambda^+_c}= 2.286$ GeV), and $\Sigma^+_c =[c(ud)_{I=1}]$ ($M_{\Sigma^+_c}= 2.453$ GeV), soon followed. One spectacular decay chain of a charmed meson produced in a neutrino interaction is given in Fig.~\ref{charmev}.

The same year, L. Lederman and collaborators, studying $p \bar p$ collisions at Fermilab, observed a new narrow state \cite{Hom1976,Herb1977,Innes1977}, the $\Upsilon$ particle. The  $\Upsilon$ is similar to the $J/\Psi$ but made by a heavier quark, soon identified with a charge $-1/3$ quark named $b$-quark ($b$ for {\it beauty}). 

At the same time that charm quark and charm spectroscopy were discovered with properties very close to what predicted, the third generation of quarks and leptons, anticipated by Kobayashi and Maskawa~\cite{KobayashiMaskawa1973} on the basis of the observed CP violation in K decays, was being unveiled. 
}
 \end{tcolorbox}
\vskip 0.3 cm

\begin{figure}[ht]
\begin{center}
\includegraphics[scale=.20]{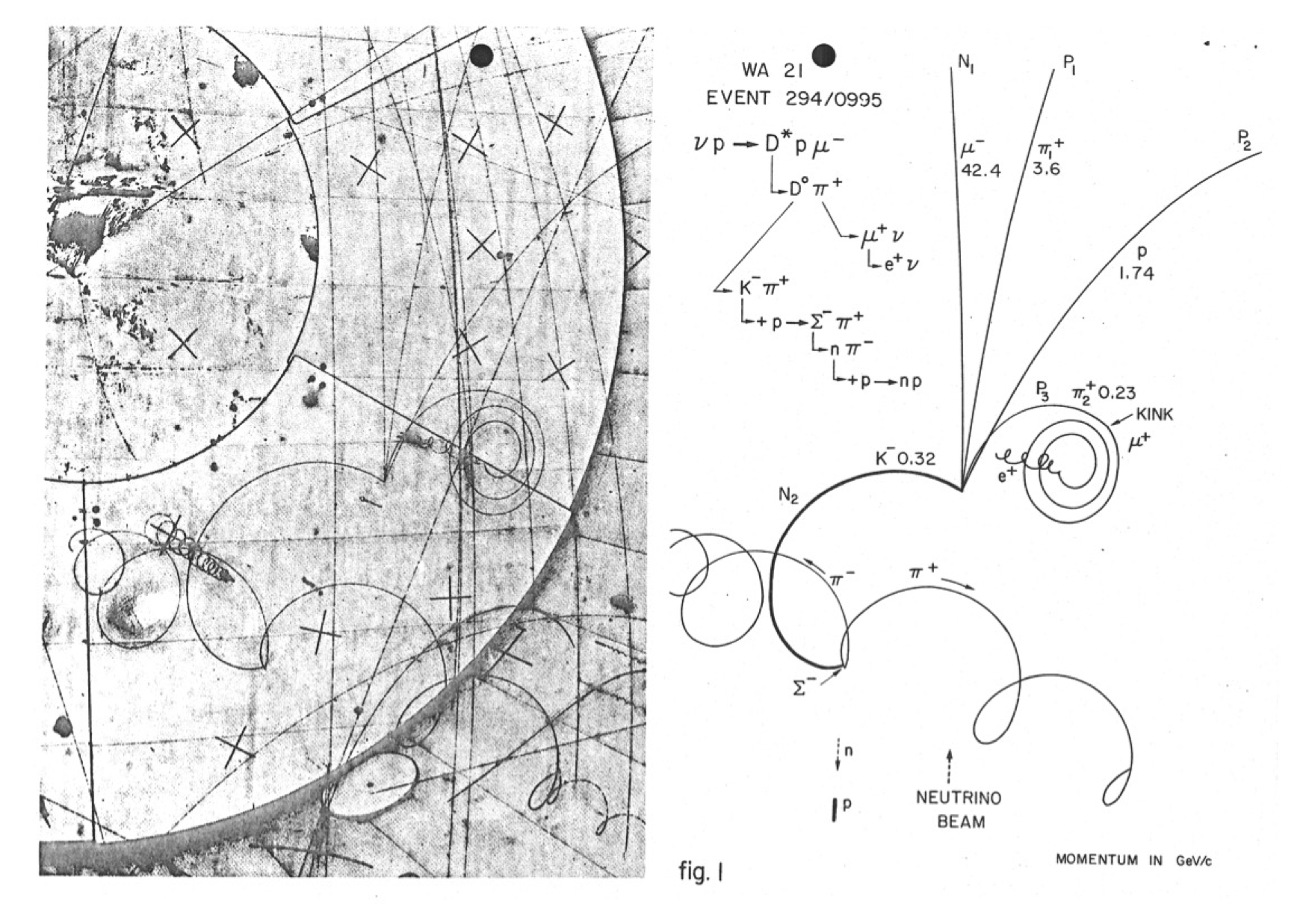}
\caption{{\footnotesize The neutrino production of an excited charmed meson, $D^\star$, is captured by this remarkable picture taken at the CERN Hydrogen Bubble Chamber, with the  decay chain of $D^\star$  fully reconstructed~\cite{Blietschau1979}. Photo on the left, explicative drawing on the right. The primary process is $\nu + p \to \mu^- + p + D^\star$ (neutrino unseen), the outgoing $p$ and $\mu^-$ correspond to traces $p_2$ and $N_1$. The primary interaction is followed by very fast decays that seem to originate also from the primary vertex: $D^{*+} \to D^0 + \pi^+,~D^0 \to  K^- + \pi^+$and give rise to the traces $N_2 (K^-)$ and $p_1, p_3$, the $\pi^+$s. The fast $\pi^+ (p_1)$ exits the chamber, but the slow $\pi^+ (p_3)$ exibits the full chain $\pi^+\to (\nu_\mu)+\mu^+\to (\nu_\mu+\nu_e)+e^+$, in parenthesis the unseen neutrinos. The negative trace $N_2$ is  identified as $K^-$ by its interaction with a proton in the bubble chamber: $K^- + p \to \Sigma^- + \pi^+$, followed by $\Sigma^- \to n + \pi^-$. The traces of $\pi^+$ and $\pi^-$ are seen to come out of the point where the $K^-$ has interacted, while the neutron goes unseen until it scatters off a proton, which is seen as the thick track below the vertex. We thank D. Haidt for calling our attention to this spectacular event. Photo CERN.}}
\label{charmev}
\end{center}
\end{figure}

 \section{Life in Rome in the 1970s.} 
 
 In Rome, Pucci and I used to see Guido and Nicola out of work, with wives and small kids. Sometime we would go to Fregene, in the nice seaside house of the Altarelli's, to Grottaferrata, in the country house of the Cabibbo's, and to the lake of Bracciano with mine and Pucci's family. We saw also other Rome professors, Giorgio Salvini, Marcello Conversi, Giorgio Careri and families. 

New younger people had joined, besides Massimo Testa and Giorgio Parisi. Keith Ellis, a young Italian-Scottish speaking student, attracted to Rome by Preparata and recruited in our group by Guido, Roberto Petronzio, and later, Guido Martinelli, also recruited by Guido. You will find their names appearing first in the literature in association with Nicola, with Guido and sometimes with me. 

From time to time the Physics Department was occupied by the students, but we could find always a quiet office in Istituto Superiore di Sanit\`a, across the road, where I worked.

Rome and Italy were struck by social turmoil and terrorism, but our was a quiet, intellectually stimulating, academic life that I remember with pleasure and that never came back.

Notably, we made a paper on the anomalous magnetic moment of the muon \cite{AltarelliEtAl1972},  the first calculation of one-loop electro-weak corrections, in competition with several distinguished colleagues \cite{JackiwWeinberg1972} \cite{BarsYoshimura1972} \cite{FujikawaEtAl1972}. 
 For the time, it was a difficult calculation and we made it at the right time. The issue is still hot, since the latest experimental determination of the muon anomaly deviates from the prediction of the full Standard Theory by $2-3$ standard deviations, corresponding to something of the same order of the electroweak corrections \cite{PDG2016}.

I moved to the University La Sapienza as full professor in 1976 and Guido took the chair shortly after, in 1980.

With John Iliopoulos back in Paris, very close relations were established between Rome and the group of Phil Meyer in Orsay. 
When Meyer's group moved from Orsay to \'Ecole Normale Sup\'erieure, in 1974, Guido Altarelli and I were visiting, living in rue d'Ulm (Keith Ellis was also around).

 The discovery of the $J/\Psi$ raised a lot of questions and we (Rome and Paris) accepted to go to Utrecht to discuss with Tini Veltman and Gerard 't Hooft, a meeting which became the annual \textit{Triangular Meeting Paris-Rome-Utrecht}, rotating among the three towns.
 
 Guido took a crucial sabbatical in ENS in 1976-1977 where he and Parisi wrote the paper on deep inelastic scattering, with the famous Altarelli-Parisi equations \cite{AltarelliParisi1977} (the most quoted French theoretical paper!). Nicola Cabibbo followed, visiting Paris VI, during my sabbatical in ENS, 1977-1978.
It was remarked, at that time, that Rome people saw CERN only from the airplane, flying to Paris. 

When in ENS, we all lived under the surveillance of Claude Bouchiat and the quiet but firm protection of Phil Meyer.

In the mid-1970/early 1980 we were busy exploring implications of the Standard Theory of Particles, as shown by the titles of our papers: neutrino reactions  \cite{AltarelliEtAl1974b}, lepton flavour non-conservation \cite{AltarelliEtAl1977a}, heavy lepton properties \cite{AltarelliEtAl1977b}, beta decay of the new quarks \cite{CabibboMaiani1978} \cite{AltarelliEtAl1982}, bounds to the Higgs boson mass \cite{CabibboEtAl1979}.

During my lectures in Gif-sur-Yvette, 1979, I came to discuss the paper by Sid Coleman and  Eric Weinberg on the higher order corrections to the Higgs boson potential. In this connection, it occurred to me that the {\it hierarchy problem} (see {\bf Box 5}) could be solved by supersymmetry \cite{Maiani1980}.

\vskip 0.3 cm
\begin{tcolorbox}[breakable, enhanced]
\vskip 0.3 cm 
\footnotesize{
{\bf Box 5. What may happen beyond the Standard Model: the Hierarchy Problem}\\
\vskip 0.1 cm
The  Standard Theory (ST) is incomplete and the missing parts imply very large mass scale, much larger than the proton mass. They are:

\indent -- Gravity: regulated by the Planck mass $M_P=(G_{Newton})^{-1/2}= 10^{19}~{\rm  GeV}$;

\indent -- Unification of the three gauge interactions of ST: Grand Unification mass $M_{GUT} = 10^{14}~{\rm  GeV}$.

 How is it possible to have the Standard Theory at the Electroweak scale, of order of $10^2$ GeV, so much smaller than the high energy scales? This is the Hierarchy Problem.
 
Spin 1/2 and 1 particles: in the zero mass limit a new symmetry is gained (chiral symmetry for spin 1/2, gauge symmetry for spin 1). This implies that higher order corrections to the mass do vanish in the limit where the bare mass vanishes. 

In the case of the electron mass, the higher order corrections are indeed of the form: 
\begin{eqnarray}
m_e(q^2)=m_0 \left[1+  C\frac{\alpha}{\pi} \log (q^2/M_{GUT})\right] \nonumber
\end{eqnarray}

with $m_0$ the bare mass and $C$ a numerical constant. The large mass is locked into the logarithm, and there is no difficulty in reproducing a mass $m<< M_{GUT}$.

In the Standard Theory, however, no increased symmetry is gained by letting the mass of the scalar field to vanish. This led different authors  \cite{Wilson1971} \cite{tHooft1980}  to declare that the Standard Theory (ST) is \textit{unnatural}.

The unnaturalness of the Higgs mass may be associated to quadratic divergences in quantum corrections and to the unnatural tuning required, between bare mass and the corrections. In contrast with the electron case, for the corrections to the  Higgs mass, one finds

\begin{eqnarray}
\mu^2=\mu_0^2+C\frac{\alpha}{\pi} \Lambda^2+\cdots \nonumber
\end{eqnarray}
with $C$ another numerical constant and $\Lambda$ a high energy cutoff. If we set, e.g. $\Lambda \sim M_{GUT}$, to obtain a physical mass $\mu\sim 100$~GeV will require an extreme conspiracy between the bare mass and the correction, their values must differ only after more than 20 digital places! 

Note the similarity with the argument presented in {\bf Box 2}; in that language, one should require an anomalously low cutoff, of order $10^3$ GeV.

A first possibility is that $C$ vanishes, due to cancellations between different contributions to the Higgs boson mass corrections. Using the analysis of Sidney Coleman and Eric Weinberg \cite{ColemanWeinberg1972} one finds that the quadratically divergent correction to the Higgs potential, arising from the exchange of particles with spin $J$, has the form

\begin{eqnarray}
V^{(2)}=\frac{1}{2}~\mu_0^2 \phi^2 +\frac{\Lambda^2}{32\pi^2}\sum _J (-1)^{2J} (2J+1)~M_J(\phi)^2 \nonumber
\end{eqnarray} 
where $M_J(\phi)\propto \phi$ is the mass that the particle acquires in the Higgs field $\phi$. Fermion and boson contributions appear with opposite sign, which means that a cancellation is possible if there is a symmetry relating particles with spin differing by 1/2 unit. A class of theories enjoying such a symmetry, called {\it Supersymmetry}, SUSY for brief, had been discovered by Wess and Zumino and by Akulov and Volkov  in the early 1970s \cite{WessZumino1974} \cite{AkulovVolkov1974} and intensively studied since then and applied to describe the electroweak unification \cite{Fayet1976} (see also \cite{Fayet2016}). 

At the end of the 1970s, several authors suggested  that the hierarchy problem could be solved by extension of the Standard Theory to SUSY, with the SUSY partners of ST particles appearing in the TeV=$10^3$ GeV range. These ideas took substance in the article by H. Georgi and S. Dimopoulos, proposing a SUSY extension of the ST with explicit SUSY breaking, see text for references.

An alternative to SUSY would be that {\it there are no elementary scalars}. The Higgs boson would be a composite of fermion fields bound by new color-like forces named {\it Technicolor} \cite{Weinberg:1975gm}~\cite{Susskind:1978ms}, manifesting themselves at a scale $ \Lambda=\Lambda_{Tech} \sim 1$~TeV.
To reproduce a Higgs boson much lighter than $\Lambda_{Tech}$, it has been assumed the Higgs to be the would-be-Goldstone boson of some global symmetry \cite{Pomarol2016} \cite{Contino2010} \cite{BellazziniCsakiSerra2014}.

The present conclusion seem to be that solving the Hierarchy problem requires new physics to appear at energies of the of order of one to several TeV.
}
\end{tcolorbox}

\vskip 0.3cm

The same idea was proposed independently by other authors \cite{Veltman1981} \cite{Witten1981a}  \cite{Witten1981b}. 

These ideas took substance in the article by Howard Georgi and Savas Dimopoulos about a supersymmetric model with explicit supersymmetry breaking \cite{DimopoulosGeorgi1981}. It was a very important work, which took many ideas about electroweak supersymmetry, formulated in particular by Pierre Fayet, now referred to as MSSM (Minimal Supersymmetric Standard Model). 

Towards the end of the 1970s, I made acquaintance with Riccardo Barbieri of Scuola Normale Superiore in Pisa. We discovered soon that we had many things in common, we both had grown up with Raoul Gatto, Riccardo and my wife Pucci had been born and grown up in Parma, and we had children of very similar ages, who soon made close friendship.

It was the beginning of a long, enduring friendship and collaboration, with work mixed with common vacations in Cortona, where Scuola Normale held every year a spring Workshop on particle physics in its beautiful Renassance villa, and on the seaside at the Elba island.

 Riccardo and I analysed Supersymmetry in the 1980s in many different contexts: the MSSM corrections to the muon g-2~\cite{Barbieri 1982}  and to the W mass renormalization~\cite{Barbieri 1983a} and the study of possible MSSM  decays of the newly discovered $W$ boson~\cite{Barbieri 1983b}, with Nicola Cabibbo and Silvano Petrarca.

\section{Three Generations for CP Violation}

After the summer conferences in 1975, it had become clear that a new heavy lepton, $\tau$ had been discovered. Due to anomaly cancellation, the simplest solution was the existence of a third generation of quarks and leptons. That prompted me, in particular after my conversation in Gif-sur-Yvette with Shelly, to come back to the issue of the CP violating phases, applied to three left-handed doublets.

In September I was in Erice, and one night I made the calculation and discovered that there was indeed one CP-violating phase remaining. 

Back to Rome, I went further to see if the CP violation implied by the phase could be consistent with what we knew about the observed CP violation in neutral $K$ decays. I could show that the prediction of the super-weak model, known to be in good agreement with data, was approximately satisfied. I was very excited and started quickly to write a short paper. While writing, it occurred to me the classification, made by L. Okun, of theories of CP violation. The theory I was considering could be classified as ``milliweak'' and as such it presented the danger of producing an electric dipole of the neutron, inconsistent with the already known, very stringent experimental upper bound.  But it was easy to prove that that neutron dipole moment had to vanish in one-loop order, so as to be naturally suppressed. Reassured by this result, I completed the paper and sent it to \textit{Physics Letters}. There was some discussion with the referee on technical issues, which made so that the paper appeared in May of the following year \cite{Maiani1976} 
But the preprint went already around before the end of 1975.
 
 Nicola was very much interested, to the point that, going to Fermilab, he made a seminar about my result.  While at Fermilab,  however, he was told  that there was a 1973 paper by Kobayashi and Maskawa with the existence of a remaining phase in the three generation model \cite{KobayashiMaskawa1973}.  
 
 At about the same time, I received a paper by Sandip Pakwasa and Hirotaka Sugawara  analysing the phenomenology of CP violation in Kaons in the Kobayashi-Maskawa framework \cite{PakwasaSugawara1976}. They had come to similar conclusions as mine, but did not consider the neutron electric dipole moment. 
 
 The one-loop result on the electric dipole was later extended by John Ellis, Mary K. Gaillard and Dimitri V. Nanopoulos who showed that it would vanish at two-loop level as well \cite{EllisEtAl1976}.
 
 The interest of Nicola in this matter remained and in Erice 1977 he transferred the CP analysis to leptons, introducing the mixing matrix for three neutrinos and analysing its effect on neutrino oscillations, CP violation included \cite{Cabibbo1978}.
 
 At present, all real angles in the neutrino mixing matrix have been measured. The experimental determination of the CP violating phase is the main challenge of the next generation of neutrino oscillation experiments.

At that time, many other contributions came out about CP violation, which was at last included in the accepted theoretical framework. 
 
 That was, in substance, the beginning of the three-generation Standard Model.

 \section{Physics at the Time of the Intermediate Vector Bosons}

At the end of the 1970s I became member of  the SPS Committee of  CERN. The SPSC was a wonderful observatory to understand the functioning  of a complicated scientific structure like CERN and I enjoyed  very much working in a mixed body of experimentalists and theorists.  I enjoyed, in particular, the physics discussions originating from experimental proposals. One of those, with my Committee fellow Mary K. Gaillard, even gave rise to an amusing work  \cite{GaillardEtAl1982}.

It was the time of CERN neutrino high energy experiments, using the SPS extracted beam, mainly proposed to achieve high precision in the determination of the Glashow-Weinberg-Salam angle, $\sin\theta$. The previous round of neutrino experiments at FermiLab had seen effects inconsistent with the Standard Theory, some  marginally, some substantially, and CERN experiments set out to clarify the issue. The round of CERN experiments eliminated all anomalies and the Standard Theory came out confirmed with flying colours.

Nicola was in the top CERN Committee, the Scientific Policy Committee. They had advised CERN to approve the construction of the proton-antiproton collider proposed by Carlo Rubbia and Simon Van Der Meer to search for the Intermediate Vector Bosons. It was a difficult machine and the positive recommendation of the SPC was a far sighted  decision.

The year 1978 I was in sabbatical in Paris and went to CERN to make a seminar, I found Carlo quite enthusiastic about the progress of the beams of the $Sp{\bar p}S$ collider, as the machine was called. 

The same year Bruno Touschek, while visiting CERN for the year had to be hospitalised in the nearby Hospital de la Tour, for a hepatic crisis. From Geneva, he was then transferred to Innsbruck, where he died on May 25.

\begin{figure}[ht]
\begin{center}
\includegraphics[scale=.50]{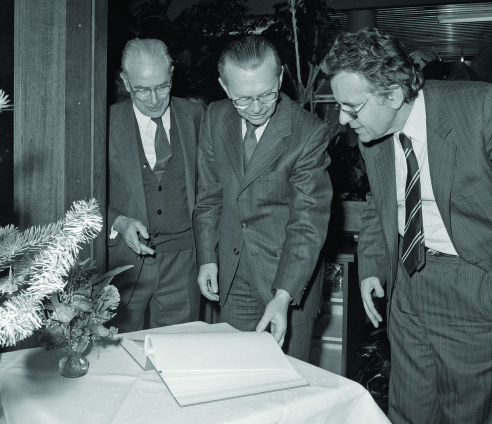}
\caption{{\footnotesize Herwig Schopper showing the Golden Book to Nicola Cabibbo (right) and Gordon Munday (left), 75th Council Session, December 1983. Photo CERN. }}
\label{schop_cab}
\end{center}
\end{figure}

In his Touschek biography, Amaldi reports~\cite[p. 2]{Amaldi 1981} that Bruno was quite disturbed to be given Room 137 in Hospital de la Tour. This was an ominous number for somebody who had spent most of life on Quantum Electrodynamics. The inverse of the fine structure constant, the number $137$, had resisted all attempts to derive it from first principles. 

Today, we look at these attempts as the sign of a happy time, when physicist were confronted by the value of one fundamental constant only, the fine structure constant. The Standard Theory enjoys two more fine structure constants (one for the weak and one for the strong interactions) and a bunch of more than fifteen Yukawa couplings, all supposed to be fundamental constants, with numerical values that range over the eleven orders of magnitude that separate the neutrino from the  Yukawa coupling of the top quark. A strong reason that there must be something {\it beyond} the Standard Theory.

The loss of Bruno was deeply felt by the physics community, particularly by us Italians, many of whom had been his students or had learned physics from him. It came, sadly, at a time when his idea of colliding rings was being successfully extended to the most difficult case of matter-antimatter collider that could be imagined.

The $Sp{\bar p}S$ started at the end of 1981 and in 1983 there were the $W$ and $Z$ bosons that stirred an unforgettable wave of excitation.

When Nicola accomplished his term, I was elected in the SPC and that was the time, 1984, when the SPC first considered the Large Hadron Collider as the next CERN large facility.

Here, I want to make a step back at the beginning of the '80s. 

In Rome, with Francesco Antonelli, Guido Corb\`o and younger collaborators, and with  Maurizio Consoli from Catania  we got interested in the electroweak corrections to the Intermediate Vector Boson properties, masses and widths. In particular, I was interested in the logarithmically dominant corrections, proportional to powers of $\alpha \log(M_W)$, that give the main pattern of the corrections \cite{AntonelliEtAl1980} \cite{AntonelliMaiani1981} \cite{AntonelliEtAl1981a} \cite{AntonelliEtAl1981b}.

We were not alone, of course, and many other people turned around these problems, notably Veltman \cite{Veltman 1977}, who pointed out that the correction to IVB masses due to the top quark is quadratic in the mass of the top while that due to the Higgs boson is logarithmic, and Marciano and Sirlin \cite{MarcianoSirlin1980}.  Competition was good to make new ideas to emerge and to check each others calculations.

However, with the electroweak interaction on the same footing as QED, people wanted to check in detail the higher order electroweak correction, which meant high precision in both theory and measurements, beyond the leading log approximation. 

The basic constants of the Standard Theory are: the Fermi constant, $G_F$, the fine structure constant, $\alpha$, and the Weinberg's angle. The first two constants could be measured with great precision, from the muon~$\beta$~decay and from low energy atomic physics, respectively. 

The third parameter, the Weinberg's angle $\theta_W$, was usually supposed to be obtained from neutrino cross sections, but I got convinced  that we would never be able to measure that to the needed accuracy. Above all because the parton model cross sections had intrinsic errors that had nothing to do with the electroweak theory and one would never get much precision out of them. Neutrino cross sections, eventually, were going to provide a test of QCD, on a different level of precision one would require from a fundamental, renormalisable theory akin to QED. 

My conclusion: forget about neutrinos, but start from the $Z^0$ and the $W$ masses, which can be measured with great precision. 

One could  {\it define} $\theta_W$ from the mass ratio:
\begin{equation}
\cos^2\theta_W=_{Def}\frac{M_W^2}{M_Z^2} \nonumber
\end{equation}
Then all other quantities in IVB physics, $M_W$, the widths, etc. could be predicted and measured with great precision\footnote{Of course one has to add genuine strong interaction corrections, determined by the QCD constant, $\alpha_S$, see e.g.~\cite{Hagiwara1994}.}. In particular, the mass of the  $W$ had to be slightly different from what predicted by the standard theory in lowest order, for that value of $\theta_W$, by a quantity that could be computed with great accuracy, from the fundamental constants and the masses of the other particles in the theory, above all the top quark and the Higgs boson. 

\begin{figure}[ht]
\begin{center}
\includegraphics[scale=.60]{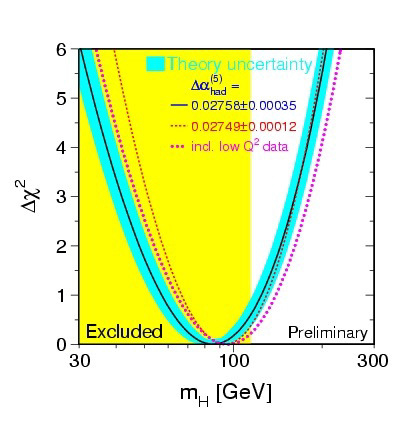}
\caption{{\footnotesize  The blue-band plot shows the constraints on the Higgs mass from precision measurements. The best fit and the width of the parabola vary, mostly due to shifts in the top mass and its uncertainty. In yellow, on the left, the region excluded by LEP experiments. Figure by the LEP ElectroWeak Working Group \cite{lepewwg}.}}
\label{bbplot}
\end{center}
\end{figure}

I made this remark in the 1982 conference in Venezia organized by Milla Baldo Ceolin \cite{Maiani1984}, after listening a long talk about the determination of $\sin\theta_W$ from neutrinos. 

After Venezia, I wrote a paper with Maurizio Consoli collecting all the formulae up to one-loop precision \cite{ConsoliEtAl1983}; a similar strategy was proposed by Z. Hioki~\cite{Hioki 1982}.

Carlo Rubbia was in Venezia and I was glad to discover later that the idea was reported in his Nobel lecture, with a figure taken from our work.

 We were at a time when the mass of the quark top was always assumed at the energy up to which the top quark had been unsuccessfully searched. In our paper we took $m_t=20$~GeV. However, Veltman's observation indicated that precision determinations of the IVB masses could give a useful hint, given the very mild dependence of IVB masses from the other unknown, the Higgs boson mass. 
 
 Eventually, this was the way we understood that the top quark mass had to be much larger -- the first observation is in \cite{FogliHaidt 1988},  see also \cite{AmaldiUEtAl 1987} -- to the point as to get to a real prediction \cite{CostaEtAl1988}, confirmed by the top quark discovery in FermiLab in 1994, with $m_t \sim 174$~GeV \cite{AbeEtAl1995} \cite{AbachiEtAl1995}.
 
 We come now to the years of LEP, which added many more electroweak observables to be compared with higher order electroweak predictions. With the top quark discovered and the precision attained at LEP, one could afford to make an overall fit to the electroweak observables, with $m_H$ the only unknown variable. This led to the famous ``Blueband Plot'', presenting the $\chi^2$ of the fit versus the Higgs boson mass and giving, at the beginning of years 2000, the upper bound $m_H \leq 200$~GeV to $90\%$ confidence level. 
 
 Well, I am now jumping over LEP, which came into operation in 1989, to our great happiness. LEP experiments have greatly improved, let me say {\it beyond expectation}, our confidence in the Standard Theory.

Four great LEP legacies are going to stay: a {\it precision test of asymptotic freedom}, with the determination of the  behaviour of $\alpha_{strong}$ up to $200$~GeV; the indication of {\it three light neutrinos}, from the $Z^0$ width; the Blueband Plot, see Fig.~\ref{bbplot}, with a very significant indication of a {\it light Higgs boson}; the determination of $e^+ e^- \to WW$, that definitely proved the existence of {\it the $WWZ$ vertex}, predicted by Yang and Mills.
 
 Few final observations about the years 1980s. 
 
\begin{figure}[ht]
\begin{center}
\includegraphics[scale=.89]{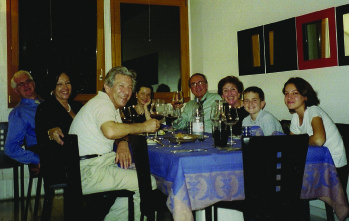}
\caption{{\footnotesize An extended GIM dinner at John's flat in Paris, circa 2000. From left: Shelly Glashow, Pucci Maiani, John Iliopoulos, Annie Iliopoulos, Luciano Maiani and Joan Glashow;  on the right, Alexandros Iliopoulos and Camilla Maiani.}}
\label{gimdinner}
\end{center}
\end{figure}
 
 In these years, we realised that the way to go to higher energies was with $p-p$ better than $p-\bar{p}$ colliders. At energies above LEP,  there are so many gluons in the proton that new particle production will be dominated by gluon-gluon fusion. It was a great simplification: you would not need quark-antiquark annihilation, as was the case at the $Sp\bar p S$, and all the gymnastic around antiproton beam cooling could be avoided.

Also, due to asymptotic freedom, we understood that a hadron machine could be almost as effective as an electron-positron machine to disentangle the basic particle reactions. So it didn't pay to go beyond LEP with electron-positron, the next machine could be proton-proton.
 
 The second observation is about a discussion that took place in mid 1980s, after Carlo Rubbia's suggestion that perhaps one could dismiss LEP construction and jump directly to make a Large Hadron Collider in the LEP tunnel. The idea met with a strong opposition in CERN and in fact it never surfaced in the outside world, as far as I know. Two reasons against the ``shortcut'' towards high energy. 
 
 Valentino Telegdi, who chaired the SPC from 1981 to 1983, strongly opposed the idea: ``What are we going to tell to our governments, that we have been planning and built a wrong machine?'' I agreed with Val. You simply could not drive CERN as you were driving a motor bike. 
 
 In addition, CERN was far from being ready for the construction of an LHC. We know with certainty, after the painful years of LHC construction, 1994-2011, how much time it took to collect the needed resources and  how much R\&D and industry preparation was needed to make the superconducting magnets and build the detectors. 
 
 Fortunately, SPC and Council, whom had been brave in taking the road to the $Sp{\bar p}S$ collider, stayed, this time, on the safe side and postponed LHC construction to LEP completion and operation.

\section{Acknowledgments}

One of us (L. M.) is grateful to Massimo Testa and Fabio Zwirner for useful comments on the manuscript and to Dieter Haidt for a very informative correspondence on the discovery of the neutral current neutrino events and for several helpful suggestions. We are also indebted to Francesco Guerra for some insightful comments. We thank the referee for very competent and constructive observations, which prompted consistent improvements of the text.

\end{document}